\documentclass[authoryear,times,12pt]{elsarticle}

\usepackage{amssymb}

\usepackage[colorlinks=true, allcolors=blue]{hyperref}
\usepackage{longtable}

\usepackage{caption} 
\usepackage[caption=false,font=footnotesize]{subfig}
\usepackage{multirow} 

\usepackage{amsfonts}       
\usepackage{nicefrac}       
\usepackage{microtype}      
\usepackage{xcolor}         
\usepackage{stmaryrd}       
\usepackage{comment}

\usepackage{bm}

\usepackage{pifont}
\newcommand{\cmark}{\textcolor{green}{\ding{52}}}%
\newcommand{\ycmark}{\textcolor{yellow}{\ding{52}}}%
\newcommand{\xmark}{\textcolor{red}{\ding{56}}}%

\begin{document}
\begin{frontmatter}



\title{Synthetic Data-Based Simulators for Recommender Systems: A Survey}


\author[inst1]{Elizaveta Stavinova}
\ead{stavinova@itmo.ru}
\author[inst2,inst3]{Alexander Grigorievskiy}\ead{alex.grigorievskiy@gmail.com}
\author[inst2]{Anna Volodkevich}\ead{volodkanna@yandex.com}
\author[inst1]{Petr~Chunaev\corref{cor1}}
\ead{chunaev@itmo.ru}
\author[inst1]{Klavdiya Bochenina}\ead{bochenina@itmo.ru}
\author[inst4]{Dmitry Bugaychenko}\ead{DmitryBugaychenko@gmail.com}

\cortext[cor1]{The corresponding author.}

\affiliation[inst1]{organization={ITMO~University},
            addressline={16~Birzhevaya~Lane},
            city={Saint~Petersburg},
            postcode={199034},
            country={Russia}}

\affiliation[inst2]{organization={Sber AI Lab},
            addressline={32/2 Kutuzovsky avenue}, 
            city={Moscow},
            postcode={121170},
            country={Russia}}
            
\affiliation[inst3]{organization={AIRI},
            city={Moscow},
            country={Russia}}

\affiliation[inst4]{organization={Sber},
            addressline={32/2 Kutuzovsky avenue}, 
            city={Moscow},
            postcode={121170}, 
            country={Russia}}

\begin{abstract}
\small
This survey aims at providing a comprehensive overview of the recent trends in the field of modeling and simulation (M\&S) of interactions between users and recommender systems and applications of the M\&S to the performance improvement of industrial recommender engines. We start with the motivation behind the development of frameworks implementing the simulations --- simulators --- and the usage of them for training and testing recommender systems of different types (including Reinforcement Learning ones). Furthermore, we provide a~new consistent classification of existing simulators based on their functionality, approbation, and industrial effectiveness and moreover make a summary of the simulators found in the research literature. Besides other things, we discuss the building blocks of simulators: methods for synthetic data (user, item, user-item responses) generation, methods for what-if experimental analysis, methods and datasets used for simulation quality evaluation (including the methods that monitor and/or close possible simulation-to-reality gaps), and methods for summarization of experimental simulation results. Finally, this survey considers emerging topics and open problems in the field.
\end{abstract}




\end{frontmatter}


\newpage

\tableofcontents

\section{Introduction}
\label{sec:definition}


A typical recommender system (RS) suggests to a user (for example, an online bookshop customer) the most suitable items (for example, books) based on available data about items (e.g. book genres), users (e.g. his/her genre preferences) and the history of interactions between them (that is usually stored as data about user--item responses e.g. scores or purchase facts). With the ever-increasing amount of data, RSs become an effective way to solve the problem of information overload \cite{Patel2017}, when a user aiming at making choice suffers from the excess of information (e.g. the number of items) available. This makes RSs a popular field of current research \cite{Lu2015,Mu2018,Rabiu2020,Wang2022,Chen2021a} and an integral part of large e-commerce systems such as Amazon, eBay, Netflix, Google, etc.

Recall that there are RS of different types, e.g. Reinforcement Learning (RL) ones, where RS is regarded as an agent, users and items are an environment, items to be recommended by an agent are regarded as actions, and user's satisfaction (estimated via clicks, views, responses, etc.) is a reward. Examples are Q-learning \cite{srivihok2005commerce}, SARSA \cite{rojanavasu2005new}, MC \cite{liebman2014dj} and others. The interaction between a user and an RS is assumed to be sequential in the case of such RSs. Other RSs include all commonly used supervised and semi-supervised approaches to the recommendation task, including collaborative filtering and content-based algorithms, see e.g. \cite{van2000using,bobadilla2011framework,vozalis2007using,luo2012incremental}.

Despite the fact that RSs have been studied for decades, their development is still a challenging task \cite{DaCosta2018,Yang2018, Ekstrand2020,Salah2020,Hug2020}. This is so due to many requirements put on modern RSs, see e.g. \cite{Slokom2018,Patki2016,Shi2019}.  Among other things, one may face the following issues within the RS development process:
\begin{itemize}
    \item the complexity, high cost and risk of training and testing RSs in the real-world environment;
    \item the insufficiency of historical data for offline training and testing RSs;
    \item the presence of privacy restrictions put on real-world data;
    \item the necessity of what-if analysis (e.g. when certain assumptions made about the expected user behavior) while training and testing RSs;
    \item the presence of bias (e.g. popularity and positivity bias\footnote{{\it Popularity bias} is the effect when of RSs tend to interact with more popular items, while {\it positivity bias} is the effect of that users of RSs rate the items they like more often, see e.g. \cite{Pradel2012,Steck2011,Huang2020}.}) in the historical data used for offline training and testing RSs, the negligence of important real-world effects in standard RS evaluation and inadequate metrics, which result in the offline-online inconsistency\footnote{The presence of the discrepancy between online and offline performance of an RS, see e.g. \cite{Huzhang2021}.}
\end{itemize}

One of the most promising approaches to cope with the above-mentioned issues is the usage of synthetic data and the modeling and simulation (M\&S) of interactions between users and RSs \cite{Ekstrand2021,Bernardi2021,Kiyohara2021,Balog2022}. It is worth mentioning here that the analysis of search requests in Google Scholar conducted in \cite{Ekstrand2021} shows that among the papers published from 2017 to 2021 and presented at world-class conferences on the subject of RSs, about 27\% of papers use synthetic data and the M\&S or discuss their applications to the task. Indeed, synthetic data and M\&S may be used for various purposes in connection to the above-mentioned issues, in particular,
\begin{itemize}
 \item to supplement and/or replace real-world data in the RS training and testing process with its synthetic analogues in the simulated environment and to overcome the  data insufficiency problem \cite{Carmen2017,Ekstrand2021} and the necessity to perform complex, costly and risky online experiments \cite{Kiyohara2021,Bernardi2021};
    \item to preserve the privacy of real-world data \cite{Slokom2018,Slokom2020arxiv,Kiyohara2021};
    \item to train and test existing and new RSs under different conditions and within different scenarios of user behavior \cite{Ie2019,Provalov2021,Balog2022};
    \item to study and control the impact of historical data bias and potential long-term  effects of interactions between users and RSs \cite{Yao2021,Huang2020}.
\end{itemize}
Luckily, nowadays there exist numerous suitable approaches for synthetic  data generation that produce data with statistically similar properties to the corresponding real-world data \cite{Patki2016,Dankar2021,Popic2019,Goncalves2020}. Before being used in the field of RSs, they have shown their effectiveness in solving various machine learning and data mining tasks, see \cite{Patki2016,Dankar2021,Popic2019,Goncalves2020}.

Below we discuss how synthetic data and M\&S can be used to cope with the above-mentioned issues of RS development in a more detailed manner.


{\it The complexity, high cost and risk of training and testing RSs in the real-world environment.}
Traditional methods of online testing of RSs\footnote{For example, in the form of Online Controlled Experiments that are also called On-Policy Evaluation in the field of RL RSs \cite{Bernardi2021,Kiyohara2021}.} require a set of responses from real users to the recommendation items provided. Usually, online users are divided into test and control groups to look for statistical differences in the effectiveness of recommendations. Although online testing seems rather suitable, it nevertheless has several disadvantages, namely, it is usually complex, time-consuming and expensive \cite{Bernardi2021,Kiyohara2021}. Moreover, it can be even risky as exploratory or incorrect online actions may be detrimental to the online user experience \cite{Huang2020}. To avoid it, one may first try to test the candidate RSs offline on historical data to select the best ones before using online tests. However, historical user data may be sparse: for each user, it contains information about the responses on a small number of items compared to the total number of items available for recommendation. This may be already resolved by the M\&S of user behavior so that one can predict the response of a user to a given item. In this way, the M\&S may particularly help to overcome the limitations of direct online tests for the candidate RSs.

{\it The insufficiency of historical data for offline training and testing RSs.} The problem of data insufficiency is described as the lack of the necessary amount of data for offline\footnote{For example, in the form of Counterfactual Policy Evaluation  that is also known as Off-Policy Evaluation in the field of RL RSs \cite{Bernardi2021,Kiyohara2021}} training an RS at the beginning of the life cycle of RSs, as well as for comparing and evaluating RSs \cite{Slokom2018,Moghaddam2019}. Actually, this is closely related to the situation described at the end of the previous paragraph. The insufficiency problem becomes especially critical for training and testing RL RSs that require the data about user-item responses that are not presented in the historical data. Indeed, such RSs may require the response for any user-item pair. To solve the problem, one may supplement the insufficient historical data by synthetic ones produced by generative user, item and user-item response models \cite{Wang2019,Balog2022}. 

{\it The presence of privacy restrictions put on real-world data.} The problem of protecting user privacy arises from the privacy rules imposed on the use of personal data (including data about user features and responses) in companies and countries. In fact, massive real-world data that can be useful for improving information systems (including RSs) is out of reach not only for public competitions and challenges, but even for R\&D departments of companies. This problem can be solved on the basis of methods for publishing company datasets without the risk of revealing confidential data \cite{Slokom2018,Dankar2021}, which have recently been actively proposed by the community of RS developers. One of the most promising approaches for it is the generation of synthetic data \cite{Slokom2018,Ekstrand2021}. Thus, instead of the publication of real-world data, a company may share a synthetic dataset itself or a pre-trained simulation framework that is able to produce realistic synthetic datasets.

{\it The necessity of what-if analysis while training and testing RSs.} Even with a sufficient amount of historical data, there are scenarios for using RSs with an inflow of new users and items with features that are qualitatively different from those in the historical set. An important issue there (which is, however,  rarely discussed in the context of RSs and synthetic data) that can influence the conclusions about the RS quality is the impact of the so-called ``No Free Lunch'' theorem for RSs. Recall that it states, roughly speaking, that all RSs produce the same quality when averaged over all possible input datasets, see e.g. \cite{Adam2019}. In that sense, a typical procedure, when the performance of an RS is evaluated on specific real-world datasets or even synthetic datasets, may not be sufficient for a comprehensive analysis of the RS \cite{Provalov2021}. Indeed, it is impossible to observe or predict the qualitative behavior of an RS by this approach if, for example, the statistical characteristics of users and/or items, functions that determine the attractiveness of items to users, as well as the modes of the RS usage by users (for example, the frequency of requests) change, or, say, data become noisier over time \cite{Provalov2021}. An explicit change of these variables under what-if analysis (scenario modelling) makes it possible to evaluate target criteria\footnote{For example, profitability or conversion metrics.} for various implementations of the RS by changing the variety of items, the preferences of users, mechanisms for providing users with information about items (communication channels, the number of displayed items, etc.) without attracting or with a limited involvement of real-world users, see e.g. \cite{Bernardi2021,Balog2022}.

{\it The presence of biases in the historical data used for offline training and testing RSs, the negligence of important real-world effects in standard RS evaluation and inadequate metrics, which result in the offline-online inconsistency.} The data which is used for training an RS is typically observational data. It means that it is collected under the influence of previously used RS. So, the data is typically biased in various ways.
For instance, \textit{popularity bias}~\cite{Huang2020} appears because users tend to interact with more popular items, hence such interactions are over-represented in the data. Another example is \textit{positivity bias}~\cite{Yao2021}. It emerges because users interact more often with items they would rate higher. Thus, the ratings present in the historical data are biased. If one ignores biases during the offline RS training and testing, they may lead to biased parameter estimation in RSs under consideration thus affecting the corresponding experimental results \cite{Huang2020}. Furthermore, this may particularly lead to offline-online inconsistency when one observes a discrepancy between the performance of the RS within offline and online testing. In a tight connection with the scenario modeling (e.g. when the bias in the historical data is modeled by using a certain parameter), parametric synthetic data may be useful in studying and controlling the impact of historical data biases on the performance of RSs \cite{ChenWang2021,Huang2020}.


 Another direction in which the M\&S may help with is accounting the relevant real-world effects. Such effects may be not taken into account under a standard accuracy-based evaluation\footnote{Using standard metrics for an RS evaluation e.g. Precision@K. See also Section~\ref{sec:C4} for more detailed discussion on metrics.} of RSs. For instance, users interact with items recommended by an RS. These interactions eventually become new training data for the RS. It is called an RS \textit{feedback loop}~\cite{mansoury_2020} and is an essential part of RS lifecycle. The M\&S and synthetic data generation allow for modeling how a particular RS performs within the feedback loop. Another effect that can be modeled is the increased (or decreased) amount of user visits. The amount of user visits essentially depends on the quality of the RS used~\cite{McInerney2021}. In this direction, the papers~\cite{Adomavicius2021,Lee2019,Zhang2020,Zhou2021,Yao2021} discuss how the M\&S can be also used to study long-term effects of interactions between users and RSs.
 
Let us also mention that the discrepancy between offline metrics and business metrics, evaluated in a real-world environment, is another reason for the offline-online inconsistency. In this context, the M\&S can be used to directly evaluate business metrics in a simulated environment or may help to develop online performance approximations. 



\medskip

In spite of the great interest towards the M\&S of interactions between users and RSs in both academic and industrial spheres, the base ideas of these methods, as well as their implementation and application methodology, seem to vary significantly  \cite{Winecoff2021}. Furthermore, as in the area of other data mining tasks, indications exist of certain reproducibility problems in today's research practice connected with the M\&S under consideration \cite{Balog2022}. In addition, the existing comparative studies of the methods and simulators for testing RSs are still few in number \cite{Ekstrand2021}. As a result, it is often impossible to reasonably determine the advantages of various modeling schemes, to learn how to properly validate models for the RSs analysis and how to adopt best practices in using such methods. The lack of methodological standards makes it difficult to reliably apply methods for modeling various scenarios of user behavior and user interactions with an RS both within the scientific research and in industrial practice.

Because of the above-mentioned aspects, new urgent tasks connected with the development and usage of methods for M\&S of interactions between users and RSs and corresponding synthetic data generation arise. The need for conduction of a survey in this direction is particularly determined by the growing requirements for increasing RSs confidence, including requirements for guarantees of the algorithms' transparency and interpretability of the results. Following this, we aim at the present study at providing a comprehensive overview of the recent trends in the field of M\&S of interactions between users and
RSs and applications of the M\&S to the performance improvement of industrial recommender engines. We confine ourselves to the preprints and papers found on the topic and published before the end of 2021. In short, the impact of this paper is as follows.

We start with the motivation behind the development of frameworks implementing the simulations --- {\it simulators} --- and the usage of them for training and testing RSs of different types (including the RL ones). Furthermore, we provide a~new consistent classification of existing simulators based on their functionality, approbation, and industrial effectiveness. Moreover, we make a summary of all the simulators we found in the research literature. Besides other things, we discuss the building blocks of simulators:
\begin{itemize}
    \item methods for synthetic data (user, item, user-item responses) generation;
    \item methods for what-if experimental analysis;
    \item methods and datasets used for simulation quality evaluation (including the methods that monitor and/or close possible simulation-to-reality gaps);
    \item methods for summarization of experimental simulation results.
\end{itemize}    

Finally, this survey considers emerging topics and open problems in the field.

\section{Simulator as a compromise and classification criteria for existing simulators}

\subsection{Simulator as a compromise between  Counterfactual Policy Evaluation and  Online Controlled Experiment}
\label{sec:CPE}

This section discusses the papers considering general problems of assessing the quality of RSs on real-world data, as well as the advantages of using synthetic data and simulators in this area.

In \cite{Bernardi2021}, the authors emphasize that industrial RSs (of different types) require an intensive development process aimed, among other things, at solving the following important tasks:
\begin{itemize}
\item identification of shortcomings and/or points of improvement of the existing RS for further development of a new version of the RS;
\item quality evaluation of the new version of the RS.
\end{itemize}
In these terms, synthetic data and simulators are well suited to address the both, namely, in simulated environments, they allow developers to run many experiments in parallel without testing in a real-world environment. At the same time, simulators are a compromise (in terms of implementation complexity, manageability and correspondence to reality) between systems for Counterfactual Policy Evaluation (offline on historical data) and systems of Online Controlled Experiments (online in the real environment). Recall that Counterfactual Policy Evaluation systems estimate the performance of a trained RS, or policy, without deploying it in the real environment or simulating the environment; it is also called Off-Policy Evaluation in the field of RL RSs \cite{Bernardi2021,Kiyohara2021}. Furthermore, Online Controlled Experiments, e.g. online A/B tests, evaluate the performance of a new RS by running it in a real production environment and testing its performance on a subset of the users of the platform; this procedure is also called On-Policy Evaluation in the field of RL RSs \cite{Bernardi2021,Kiyohara2021}.

The authors note that the applicability of Counterfactual Policy Evaluation systems is limited because a new counterfactual model must be built for each individual study; experiments within Online Controlled Experiments systems are costly and do not allow manipulation of variables such as user preferences or market conditions.

In the same direction, in order to refute the counter-claim about the preference for experiments using only real-world data within Counterfactual Policy Evaluation (or Off-Policy Evaluation in the context of offline RL), the authors of \cite{Kiyohara2021} describe the main risks of such experiments and, in particular, point out the problem of their reproducibility. This problem is related to the lack of sufficient publicly available datasets for such experiments, which is caused, first of all, by the need to ensure the reliability of the collection and confidentiality of personal data. The latter, in practice, is associated with significant financial and production costs.
What is more, it is also noted that publishing real-world data to test offline RL RSs is complicated by the fact that new policy evaluation requires access to the environment. Thus, it is extremely difficult to conduct a reliable and comprehensive experiment using only real-world data --- this is the bottleneck of traditional methodologies for the development of RL RSs using offline RL and Counterfactual Policy Evaluation (Off-Policy Evaluation) in practice.

Having confirmed with arguments the high importance of using simulators and synthetic data for the development of RSs,  the authors of \cite{Bernardi2021} propose a set of general principles for constructing industrial simulators:
\begin{itemize}
\item  the implementation of the RS should be completely independent of where it is used, in a simulated or in a real-world environment;
\item users of a simulator should specify only the assumptions for the experiments, while all other parameters should, as far as possible, be derived from real-world data (e.g. by the way of  generating synthetic data by generative models trained on real-world data);
\item a simulator should allow its users to efficiently develop simulated environments with reusable and extensible components by allowing developers to manipulate variables and explicitly make assumptions and interventions (say, within what-if analysis in the form of a parametric scenario modeling).
\end{itemize}

Partly using the above-mentioned principles and having in mind the variety of existing simulators, we further propose several criteria for their classification.

\subsection{Proposed classification of simulators}

 There is no generally accepted approach to the simulators classification, but a set of criteria for simulators' comparison was proposed in \cite{shi2019pyrecgym} and revised in \cite{Bernardi2021}. The criteria from the two above-mentioned sources include: 
 \begin{itemize}
    \item customization for RSs;
    \item recommender task and real-word environment specifics, e.g. User Feedback Flexibility, Generalized Recommendation Task, Marketplace Simulation;
    \item nature of the dataset or ability to work with external real/benchmark datasets;
    \item flexibility and customizability of the environment, namely, Modular Environments Support proposed in \cite{Bernardi2021}.
\end{itemize}

 We revised and expanded those criteria to include the aspects important for the practitioners aiming to use an existing simulator or develop a new one. In what follows, we distinguish simulators with respect to their:
\begin{itemize}
    \item {\it functionality} in the sense of the composition and properties of the included simulator's functional components that determine the simulation pipeline;
    \item {\it approbation} in the sense of reproducibility of the experimental study conducted with the simulator;
    \item industry {\it effectiveness} in the sense of its suitability for industrial deployment.
\end{itemize}

We define and substantiate the simulator's functional components and comparison criteria in Sections \ref{sec:functionality}--\ref{sec:industry_eff} and perform the evaluation of existing simulators by means of the criteria in Sections \ref{func_review}--\ref{sec:ind_eff_review}. We also study various purposes of the simulators' development and present the results in Section \ref{sec:goal}. 

\subsection{Simulator's functionality}
\label{sec:functionality}


\begin{figure}
    \centering
    \includegraphics[width=\linewidth]{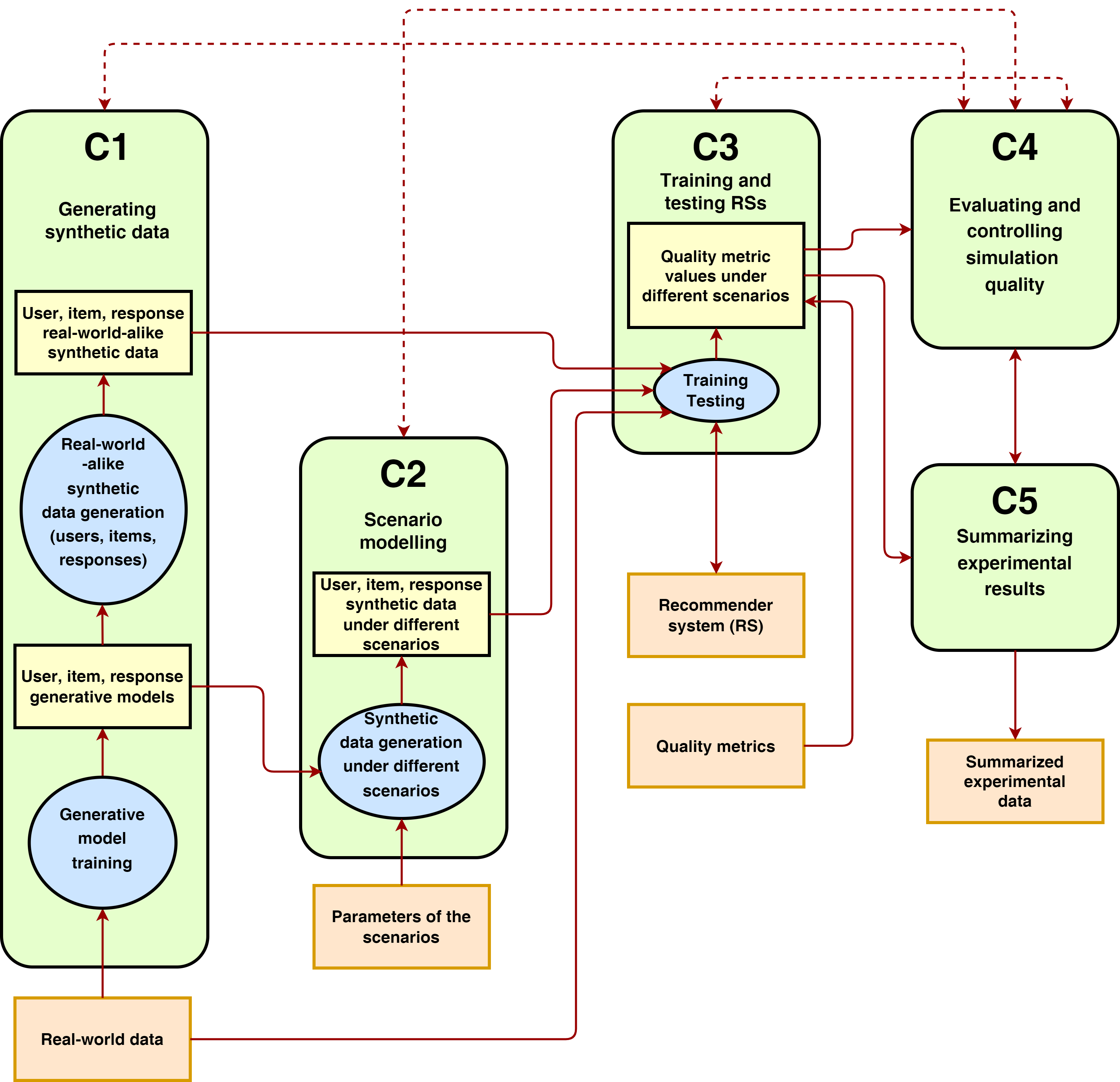}
    \caption{The proposed simple component scheme of a generalized simulator. For simplicity, it is supposed that all the synthetic/real-world data necessary for training and testing the RS are sent to {\sf C3} (from {\sf C1} and/or {\sf C2}). In general, the RS may directly use the generative models/synthetic data from {\sf C1} and/or {\sf C2} and the real-world data on request within the training and testing process.
    }
    \label{fig:component-scheme}
\end{figure}

We asses the simulator's functionality by the presence of the functionality components from the generalized simulator scheme proposed by us and shown in Figure~\ref{fig:component-scheme}:
\begin{itemize}
\item {\sf C1}: a component for training user, item and user-item response models on real-world data and generating synthetic data that are similar to the real-world one in a certain sense; 
\item {\sf C2}: a component for scenario modeling and what-if analysis that aims at generating synthetic data for further examining and evaluating possible events or scenarios that could take place in the interactions between users and RSs under chosen assumptions about users, items and user-item responses;
\item {\sf C3}: a component for training and/or testing RSs (possibly of different types) on synthetic and/or real-world data about users, items and user-item responses;
\item {\sf C4}: a component for evaluating and controlling the simulation quality that monitors and/or closes simulation-to-reality gaps and inconsistencies within different simulator's components and/or monitors and overcomes data biases and long-term negative effects of interactions between users and RSs within the simulation;
\item {\sf C5}: a component for summarizing and/or analyzing (e.g. visualization of) extensive experimental results (e.g. for a wide range of scenario parameters) about training and/or testing RSs on synthetic and/or real-world data.
\end{itemize}

Note that a simulator as a set of connected components in general and {\sf C1}, {\sf C2} and {\sf C3} in particular follows the principles in \cite{Bernardi2021}, while the motivation to include {\sf C4} and {\sf C5} stems from the necessity to evaluate and control the quality of simulation at different stages and to present results of extensive experiments in a reasonable form.

We consider the components {\sf C1} and {\sf C3} as basic for what we mean by a simulator (i.e. a framework implementing synthetic data-based simulations of interactions
between users and RS). The input of {\sf C1} is the real-world data, that is further pre-processed and used for synthetic data generation models training. These models include models for generating user and item profiles (optional) and models for generating synthetic user-item responses. It is possible that some real-world-alike synthetic data are generated in {\sf C1} and further used in {\sf C3}. From the other side, the corresponding generative models may be the output of {\sf C1} that are further used as an input for {\sf C2} where synthetic datasets are generated under different scenarios (their parameters are input).

Note that {\sf C2} may be omitted in some simulators, in that case, the synthetic data from {\sf C1} goes directly into {\sf C3}. Also, some intermediate cases with the usage of scenario modeling, synthetic data and real-world data are possible. RSs training and testing is performed in {\sf C3} under a certain choice of RSs and quality metrics provided by the simulator's user. Recall that RSs may be of different types and therefor a repeated interaction of the RS and the simulator's component {\sf C3} is possible. The metric values from C3 can be used by {\sf C4} and {\sf C5}.

Moreover, in {\sf C4} (optional) the simulation quality evaluation and control may be carried out in different senses (estimation of biases, simulation-to-reality gap and/or negative effects of interactions between users and RSs in the simulation). It is possible that, depending on the results in {\sf C4}, some other components may be tuned somehow (for example, models in {\sf C1} may be updated or re-trained to produce synthetic data of better quality).

Finally, {\sf C5} uses as input the results from {\sf C3}  and/or {\sf C4} to produce a summarization/generalization of extensive simulation experiments (e.g. different sorts of visualization) that can be used by the simulator's user.

In what follows, we will consider particular implementation of the above-mentioned components in existing simulators in Section~\ref{sec:variants}.

\subsection{Simulator's approbation}\label{approbation_def}

This block of criteria is motivated by modern requirements on the reproducibility of research in the field of RSs development. In particular, we include these criteria in order to follow the general line of \cite{Dacrema2021,Winecoff2021} about the need for careful analysis of the interpretability and reproducibility in the field. In this regard, we distinguish simulators by that they have:
\begin{itemize}
    \item been tested on {\it open} real-world/synthetic data and the results of that are presented;
    \item an open source code available;
    \item a detailed documentation available;
    \item been applied for training/testing RSs of different types (RL and not RL).
\end{itemize}

Let us mention that the criteria related to the present of open source code and a detailed documentation may be made more strict from the industry view point. For example, one may ask if the open source code is under ongoing development and has a  live community around it, or if the documentation contains quick start guides to quickly understand simulators’ functional and speed up its’ deployment. Criteria of this kind are introduced in Section~\ref{sec:industry_eff}.

\subsection{Simulator's industry effectiveness}
\label{sec:industry_eff}

We propose a set of industrial criteria based on the resent publication \cite{Bernardi2021,McInerney2021,Gauci2018} and the requirements we developed based on our own industry experience in RSs for a financial institution with hundreds of thousands of clients.

The main criterion to select a simulator for industry usage is applicability to the business area and recommendation task, which is studied in Section \ref{func_review}.  The others criteria are related explicitly to industrial usage, cover industrial results, quality or maturity of the developed solution.  

Real production RSs are characterized by a large data volumes and low latency. Therefore a simulator should be able to work with large datasets and produce necessary amount of responses to meet simulation goals. It also should bring a business-value which could be measured as a better consistency between simulator and reality or business metric growth. To evaluate this, we propose Industry applicability criteria.
On the other side an open source simulator should be easy to use and allow to extend its functionality, which could be named as Solution maturity criteria.

\subsubsection{Industry applicability criteria}

\textit{Large data volumes and technology stack}
\begin{itemize}
\item  Scalability. Industrial recommendation systems often work with large volumes of data and a simulator should be able to process and generate necessary data volumes \cite{Gauci2018}. This could be achieved, for example with specific modules for big data preprocessing, availability of hardware acceleration or implementation of batch inference to speed up response generation.
 \item Technology stack. The information about programming language and additional frameworks used by the simulator, e.g. those, which enables GPU acceleration or cluster data processing is important to understand a simulator's applicability for a particular practitioner's needs and available computational infrastructure.
\item Experiments on large datasets. It is important for industrial usage to make sure a simulator could generate the amount of data comparable to the real and process it within a reasonable time \cite{Gauci2018}. Thus presence of experiment and information about dataset sizes could be indirect evidence of industrial applicability.
\end{itemize}
\textit{Focus on industry tasks}
\begin{itemize}
\item Industrial partner. Some simulators are developed by or in a cooperation with industry practitioners, which could increase industrial applicability. A stronger signal is a simulator's usage by the industry companies other than the main industrial partner.
 \item Simulation and real-environment results consistency. Offline-online inconsistency is a known challenge for recommender and search systems and simulators for those systems \cite{Gao2021}. Thus we consider consistency between recommender results in simulated environment and A/B tests results is a strong signal of a simulator quality. 
 \item Financial effect. Some simulators not only improve RSs quality, but could bring financial effect, e.g. \cite{Shi2019}, which in our opinion also indicates industrial applicability.
\end{itemize}

\subsubsection{Solution maturity criteria}
\begin{itemize}
 \item Entry threshold. Presence of detailed documentation and quick start guides allows to quickly understand simulators' functional and speed up its' deployment.  In contrast to criterion from Section \ref{approbation_def} here we draw attention to the sufficiency of documentation and examples for easy implementation and further development. 
 \item Open source code with ongoing development and live community. Presence of a source code is considered in Section \ref{approbation_def}, but it is not sufficient for practitioners, who want to use a simulator for a long time. Continuing development after publication, regular commits or releases, issues resolving and learning materials creation shows simulators’ demand from industry and research.
\item  Modular structure and customizability.  A Simulator may contain a lot of components, e.g. User Preference Model, Item Availability Model, Delay Model \cite{Bernardi2021}, User Visit Model \cite{McInerney2021} and other models to better approximate the real-world environment. An ability to add custom models and reuse implemented models in different environments make a simulator more suitable for industrial usage. This criterion may be formulated as a Modular Design of a simulator and also mentioned in \cite{Bernardi2021}.
\end{itemize}

\section{The review and classification of existing simulators}\label{sec:eval}

\subsection{Brief description of simulators}\label{sim_overview} 

We start this section with a brief description of simulators that shows the versatility of approaches to their development and applications. Note that descriptions of all the simulators under consideration in a short form are given in Table~\ref{tab:existing_sims} in Appendix.

 RecSim \cite{Ie2019} is a customizable platform for modeling sequential user interactions with RL-based RSs. RecSim provides flexibility in the creation of a modeling environment: for example, it is possible to determine some aspects of user behavior (e.g. user preferences on certain properties of items can change over time), as well as the attribute structures of items offered to users. Note that the framework implementation in the case of modeling a complex system can be complicated because the approach assumes the creation of an abstraction levels platform for modeling certain aspects of user behavior. The positive point is that the authors of RecSim assume the re-usage of the simulator's components in different simulation environments.

In \cite{Huang2020}, SOFA (Simulator for OFfline leArning and evaluation), which is based on a method for correction of the interaction bias present in the data, is proposed. The problem of interaction bias is often ignored in simulators, where modeling of user feedback based on historical data is usually used. It is not possible to avoid the usage of historical data in the process of testing RL-based methods, as the online testing process can lead to an unfavorable user experience. A method, that corrects interaction bias present in historical data before this data will be used to construct user behavior models, is proposed in \cite{Huang2020} as a solution to the problem of interaction bias. In addition, the authors present a new method to evaluate the impact of interaction bias on the quality of the recommender algorithm. Both methods (bias correction and evaluation of bias impact) are the basis of the proposed simulator.

Furthermore, the simulator YHT\footnote{Here and below, for the simulators YHT \cite{Yao2021} and ZZA \cite{Zhou2021} we compile the simulators' names using the first letters of the first three author surnames as no names are proposed in the original papers.} \cite{Yao2021} is developed to analyze the effects of RSs performance in the case of different types of user behavior. This simulator allows (a) to analyze the influence of the RS on the dynamics of user preferences and (b) to explore how the system performs in the case not only of an ``average user'' but also in extreme conditions of atypical user behavior. The proposed simulator allows analysis of the relationship between user preferences and the recommended items in order to better understand the effects of popularity bias that occur over time. In addition, the authors develop special user models to infer the long-term impact of the RS, as well as its ability to respond to explicit and implicit user preferences.

As an alternative to the standard training of RL-based algorithms on historical data,Virtual-Taobao \cite{Shi2019}, where modeling is carried out on the basis of historical data of users' behavior, is proposed. Virtual-Taobao allows training of an RL-based model without the expenses connected with the testing in the real world. To increase the simulation accuracy, Virtual-Taobao includes the following components: GAN-SD (GAN-for-Simulating-Distribution) for the generation of users' attributes with a distribution similar to the original one and MAIL (Multi-agent Adversarial Imitation Learning) for the generation of users' actions. To avoid the simulator overfitting, an ANC (Action Norm Constraint) is proposed to regularize the strategy of the RL-based algorithm.

Let us note a few more works that are rather close to the considered subject, but cannot be regarded as simulators. However, the tools developed there for testing of RSs and other systems can be used as components of simulators. For example, frameworks for determining the RL strategy by the data are proposed in \cite{Ho2016, Yu2019}. In \cite{Chen2021, Bai2019}, frameworks for training RL-based RSs are proposed. However, in these works simulation only of some RL system components is considered. For instance, simulation of user behavioral preferences is proposed in \cite{Chen2021}, simulation of user and RL agent interactions is regarded in \cite{Bai2019}. These frameworks are not included in tables, as they are not focused on the RS training as well as on the analysis of RSs quality (i.e. do not contain the component {\sf C3}  in a sense, see Section~\ref{sec:functionality}). 

Moreover, there are three works that are aimed at training and testing RL RSs. The first one is ReAgent \cite{Gauci2018}, a platform for applied RL which allows the implementation of RL models on industry problems with large datasets. The second one is MARS-Gym \cite{Santana2020}, a framework for building and evaluating RL agents for recommendations in marketplaces. The third one is Open Bandit Pipeline \cite{Saito2020}, a framework for bandit algorithm comparison and off-policy evaluation. Since these two frameworks do not contain {\sf C1}, they are not regarded as simulators in our work (see Section~\ref{sec:functionality}). (Open Bandit Pipeline has an option to generate synthetic data, but this data is simple and does not have a clear description of its generation.)

Let us also mention two simulators that should be mentioned here as their general construction schemes are similar to those on Fig.~\ref{fig:component-scheme}, but they are created not for RSs but for search engines training and testing. The first simulator is AESim \cite{Gao2021}, which has such components as a virtual user module and feedback module (generation of synthetic user profile and queries, {\sf C1}), ranker system (analogue of {\sf C3}). A disadvantage of AESim is the lack of reproducibility: the authors do not test their simulator on open data and do not provide the source code. The second one is RerankSim \cite{Huzhang2021}, a simulator from Alibaba for training and testing ranking systems. It also has {\sf C1}, analogue of {\sf C3} for ranking algorithms, moreover, it has {\sf C4} in a form of a component for solving the problem of offline-online inconsistency. The further analysis of solutions for such a problem proposed in \cite{Huzhang2021} is carried in Section~\ref{sec:C4}. What is more, RerankSim is tested on open data and the corresponding code is available online (\href{https://github.com/gaoyq95/RerankSim}{Link}).

Moreover, in \cite{Kiyohara2021} a potential usage of simulation for offline training and testing of RL RSs is analyzed. The authors provide a scheme of their simulation platform and its workflow, but this project is only at the beginning now, as it does not have a particular implementation of the proposed scheme to the best of our knowledge.

\subsection{Review of simulation goals}
\label{sec:goal}

In Section \ref{sec:definition} we listed some problems in RS development and a brief overview of how simulation and synthetic data generation help to solve them. We believe that the purpose of creation or specialization is an important characteristic for a practitioner, looking for a simulator for a specific recommendation task and business area. We analyzed the purpose of simulators' creation and the research goals of related publications. The close approximation of real environment and accurate estimate of online RS performance could be stated as fundamental goals of each simulator and is discussed in Sections \ref{sec:C1_imp}--\ref{sec:C4}. More practical goals, which often become a motivation for research studies and frameworks development, are listed below:
\begin{itemize}
    \item Synthetic data generation for specific recommender task or environment. This includes anonymous data generation, data enrichment or generation of closer-to-reality task-specific data (e.g. enhancement of the environment with a User Visit Model or simulating additional agents, such as content creators or vendors). Related research papers evaluate synthetic data quality with statistical tests on historical data, or conduct a comparison of RS performance on synthetic and real data.
        \item What-if analysis and long-term effects evaluation. Researchers evaluate RS performance under different conditions and study the long-term effects of RS in a simulated environment.
    \item Training and evaluation of RL-based RS, especially on-policy RL models. The main success factor here is a better quality of the model, trained in the simulator, compared to the previous RS.
\end{itemize}

Table \ref{tab:purpose} groups simulators by the main goal and provides notable details on specialization, e.g. recommendation task or business area. We do not consider the RS evaluation as an individual goal in the table as it is often combined with the other goals.

It could be noticed that simulators for RL-based RS training and evaluation are becoming popular in recent years. Some simulators aim to study RS and provide a ready-to-use solution for the practitioners. A significant amount of simulators are specialized for a business area or RS specifics. 

\begin{table}
\caption{Goals of simulators and relevant research studies}
\label{tab:purpose}
\begin{tabular}{p{3cm}|p{10.5cm}}
Simulator's/study goal & Simulators \\ 
 \hline
Synthetic data generation with respect to specific recommender task or environment & 
DataGenCARS \cite{Carmen2017}, CARS-specific data \newline 
GIDS \cite{Jakomin2018}, Inter-dependent Data Streams \newline 
SynRec \cite{Slokom2018,Slokom2020arxiv}, synthetic data generation for privacy protection \newline
CF-SFL \cite{Wang2019}, CF performance boost with synthetic data \newline
Accordion \cite{McInerney2021}, modeling of user visit \\
\hline
What-if analysis and long-term effects evaluation & 
SIREN \cite{bountouridis2019siren} long-term effects evaluation in online News environments \newline
RecLab \cite{krauth2020offline}, study of offline-online metrics relation and the effects of exploration  \newline
T-RECS \cite{Lucherini2021}, long-term effects evaluation for multistakeholder problems \newline
SynEvaRec  \cite{Provalov2021}, evaluation of RSs under different scenarios of user behavior \newline
YHT \cite{Yao2021}, evaluation of RS under different types of user behavior \newline 
ZZA \cite{Zhou2021}, longitudinal impact of preference biases on RS performance
\\
\hline
RL-based RS training and RS evaluation &
RecoGym \cite{Rohde2018}, product Recommendation in Online Advertising   \newline
CLL \cite{Chen2019}, 
RecSim \cite{Ie2019}, 
UserSim \cite{Zhao2021}, \newline
Virtual-Taobao \cite{Shi2019}, AESim \cite{Gao2021}, RerankSim \cite{Huzhang2021}, Online Retail Environment \newline 
SOFA \cite{Huang2020}, debiasing of user-item rating matrix before building user response model  \newline
RecSim NG \cite{Mladenov2020}, multi-turn and multi-agent RS in ecosystem environment
\end{tabular}
\end{table}

\subsection{Simulator's functionality review}
\label{func_review}

Table~\ref{tab:functional_components} summarizes the properties of simulators by means of functionality\footnote{Note that the presence of components is assessed based on the simulator's descriptions in the papers but not on the source code.}. It is seen that the table shows a variety of combinations of functional components used in the simulators. Recall that components {\sf C1} (synthetic data generation) and {\sf C3} (training and testing RSs) are the determining components of a simulator. They are presented (at least implicitly) at every simulator from the table. The situation with the rest of the components is different. Indeed, component {\sf C2} (scenario modeling) is available for about a half of the simulators (8 of 18), while for the two of them it is implicit and thus has no particular implementation. This direction seems to be important for future development. The problem of simulator quality control, bias and other negative effects has not been considered before 2019, and only about a fourth of all simulators (4 of 18) contain the corresponding component {\sf C4} (evaluation and control of simulation quality). In general, it seems that there is no universally recognized methodology in this direction and therefore one can see a diversity in {\sf C4} implementation options. While the component {\sf C5} (summarization of experimental results) seems to be necessary for presenting simulation results in a reasonable form, about a half of the simulators in the table (10 of 18) do not provide such tools.

It is however important to mention that the presence of components is mainly defined by the simulator's goal, see Section~\ref{sec:goal}.

\begin{table}
\caption{Simulator's functionality features (ordered by the year of publication and alphabetically within a year). The mark \cmark\, corresponds to the presence of a component in the simulator's structure, while the mark \xmark\, to the absence of the component. The mark \ycmark\, means that the component is implicitly present (it is assumed that the component is needed but any variant of its development is absent) or it is in a basic form.}
    \centering
    \begin{tabular}{ l|c|c|c|c|c} 
 &    \multicolumn{5}{c}{Functional component}\\
\cline{2-6}
\raisebox{1.5ex}[0cm][0cm]{Simulator}  & {\sf C1} & {\sf C2} & {\sf C3} & {\sf C4} & {\sf C5}  \\ 
 \hline
DataGenCARS \cite{Carmen2017} & \cmark & \xmark & \ycmark & \xmark & \cmark  \\
GIDS \cite{Jakomin2018} & \cmark & \cmark & \ycmark & \xmark & \cmark  \\
RecoGym \cite{Rohde2018} & \ycmark & \xmark & \cmark & \xmark & \xmark  \\
SynRec \cite{Slokom2018} & \cmark & \xmark & \ycmark & \xmark & \xmark  \\
SIREN \cite{bountouridis2019siren} & \cmark & \cmark & \cmark & \xmark & \cmark \\
CLL \cite{Chen2019} & \cmark & \xmark & \cmark & \xmark & \cmark \\
RecSim \cite{Ie2019} & \ycmark & \ycmark & \cmark & \xmark & \xmark \\
Virtual-Taobao \cite{Shi2019} & \cmark & \xmark & \cmark & \cmark & \cmark  \\
CF-SFL \cite{Wang2019} & \cmark & \xmark & \cmark & \xmark & \cmark \\
SOFA \cite{Huang2020} & \cmark & \xmark & \cmark & \cmark & \xmark  \\
RecLab \cite{krauth2020offline} & \cmark & \cmark & \cmark & \xmark & \cmark \\
RecSim NG \cite{Mladenov2020} & \ycmark & \ycmark & \cmark & \xmark & \xmark  \\
T-RECS \cite{Lucherini2021} & \ycmark & \xmark & \cmark & \xmark & \cmark  \\
Accordion \cite{McInerney2021} & \cmark & \xmark & \ycmark & \xmark & \xmark  \\
SynEvaRec \cite{Provalov2021} & \cmark & \cmark & \cmark & \xmark & \cmark  \\
YHT \cite{Yao2021} & \cmark & \cmark & \cmark & \ycmark & \xmark  \\
UserSim \cite{Zhao2021} & \cmark & \xmark & \cmark & \xmark & \xmark 
\\
ZZA \cite{Zhou2021} & \cmark & \cmark & \cmark & \ycmark & \cmark  \\
\end{tabular}
    \label{tab:functional_components}
\end{table}

\subsection{Simulator's approbation review}

Table~\ref{tab:approbation} summarizes the properties of simulators by means of approbation. We also provide a GitHub link to the simulator if available.

One can notice that simulators are applied for training and testing RSs of different types (RL and not RL). In both cases, however, it happens that open real-world or synthetic datasets\footnote{Note that we collect the open real-world datasets that can be used for training and testing RSs of the both types in Table~\ref{tab:datasets}.} used for that and the corresponding source code are unavailable.  Furthermore, even if the source code is available, one may encounter significant difficulties in its use due to the lack of proper documentation. These facts make the problem of reproducibility essential in the field.

Generally speaking, even the presence of open well-documented source code, datasets and results does not guarantee complete interpretability, generalizability, and replicability of the study. Such cases have been reported e.g. in \cite{Dacrema2021,Winecoff2021}. For example, \cite{Winecoff2021} discusses four papers on the study of RS filter bubble effects which came to conflicting conclusions on the effect mechanism. It turns out that the studies have not agreed upon scientific terms and this led to heterogeneous and sometimes contradictory results. Based on the example, the authors of \cite{Winecoff2021} argue that the overall situation in the field results in a ``patchwork of theoretical motivations, approaches, and implementations that are difficult to reconcile.'' Furthermore, the ``lack of consensus on best practices in simulation studies of RS has limited their potential scientific impact''.

From our side, let us also mention that the extensive experimental comparison (under the same settings such as datasets, quality metrics and RSs under consideration) of existing simulators is yet to be performed according to our review of papers about simulators. Consequently, it is still an open question which simulators can be called state-of-the-art and which one is preferable for training and testing RSs on a synthetic user, item and user-item response data. The only exception we could find is UserSim \cite{Zhao2021} which contains results of several experiments in this direction. Indeed, it compares user behavior models contained in some simulators from Table~\ref{tab:functional_components} but nevertheless, it does not compare the corresponding simulators as whole pipelines.

\begin{table}
\caption{Simulator's approbation features (ordered by the year of publication and alphabetically within a year). The mark \cmark\, corresponds to the presence of the approbation feature in a paper or in a source code, while the mark \xmark\,  to the absence of that. The mark \ycmark corresponds to the presence of open-source code that can be used to reproduce the results of certain experiments on a particular dataset but not on an arbitrary one.}
    \centering
    \begin{tabular}{ l|c|c|c|c|c} 
Simulator & \rotatebox[origin=c]{90}{\begin{tabular}{@{}c@{}}Tested on open data\\+the results presented\end{tabular}}   &  \rotatebox[origin=c]{90}{\begin{tabular}{@{}c@{}}Detailed documentation\end{tabular}} & \rotatebox[origin=c]{90}{Open source code} & \rotatebox[origin=c]{90}{Applied for RL RSs} & \rotatebox[origin=c]{90}{Applied for not RL RSs}\\ 
 \hline
DataGenCARS \cite{Carmen2017} & \ycmark  &  \xmark & \xmark & \xmark & \cmark\\
GIDS \cite{Jakomin2018} & \cmark  &  \xmark & \xmark & \xmark & \cmark\\
RecoGym \cite{Rohde2018} & \xmark &  \cmark & \cmark  \href{https://github.com/criteo-research/reco-gym}{Link} & \cmark & \xmark \\
SynRec \cite{Slokom2018} & \cmark  &  \xmark & \cmark  \href{https://github.com/SlokomManel/SynRec}{Link} & \xmark & \cmark \\
SIREN \cite{bountouridis2019siren} & \xmark  & \cmark & \cmark 
\href{https://github.com/dbountouridis/siren}{Link} & \xmark & \cmark\\
CLL \cite{Chen2019} & \cmark  &  \xmark & \cmark  \href{https://github.com/xinshi-chen/GenerativeAdversarialUserModel}{Link} & \cmark & \xmark\\
RecSim \cite{Ie2019} & \xmark  &  \cmark & \cmark  \href{https://github.com/google-research/recsim}{Link} & \cmark & \xmark\\
Virtual-Taobao \cite{Shi2019} & \xmark  &  \xmark & \ycmark  \href{https://github.com/eyounx/VirtualTaobao}{Link} & \cmark & \xmark\\
CF-SFL \cite{Wang2019} & \cmark  &  \xmark & \xmark & \xmark & \cmark\\
SOFA \cite{Huang2020} & \cmark  &  \xmark & \cmark  \href{https://github.com/BetsyHJ/SOFA}{Link} & \cmark & \xmark\\
RecLab \cite{krauth2020offline} & \cmark  & \cmark & \cmark 
\href{https://github.com/berkeley-reclab/RecLab}{Link} & \xmark & \cmark\\
RecSim NG \cite{Mladenov2020} & \xmark  &  \cmark & \cmark  \href{https://github.com/google-research/recsim_ng}{Link} & \cmark & \xmark\\
T-RECS \cite{Lucherini2021} & \xmark  &  \cmark & \cmark  \href{https://github.com/elucherini/t-recs}{Link} & \xmark & \cmark\\
Accordion \cite{McInerney2021} & \cmark  &  \xmark & \cmark  \href{https://github.com/jamesmcinerney/accordion}{Link} & \xmark & \xmark\\
SynEvaRec \cite{Provalov2021} & \cmark  &  \xmark & \cmark  \href{https://github.com/vldpro/SynEvaRec}{Link} & \xmark & \cmark\\
YHT \cite{Yao2021} & \cmark  &  \xmark & \xmark & \xmark & \cmark\\
UserSim \cite{Zhao2021} & \cmark  &  \xmark & \xmark & \cmark & \xmark\\
ZZA \cite{Zhou2021} & \cmark  &  \xmark & \xmark & \xmark & \cmark\\
\end{tabular}
    
    \label{tab:approbation}
\end{table}

\subsection{Simulator's industry effectiveness review}\label{sec:ind_eff_review}

Some researchers clearly state the industrial advantages of the developed simulators, others do not, thus we will consider a simulator as meeting a criterion if those are clearly stated in a related research paper/documentation or could be concluded from an available source code. Some simulators mentioned in this section are not considered in other sections as they do not meet a simulator definition from Section \ref{sim_overview}. Despite this, those simulators have noticeable industry advantages and are thus considered here.

Some simulators do not have available source code and thus could not be directly reused by other practitioners (See the source code availability in Table~\ref{tab:approbation}). The others have not been updated after the related research paper publication which increases the risk of simulator obsolescence and incompatibility with new versions of the programming language and required packages. Those still could be interesting for industry practitioners as an experimental platform and a source of methods and techniques. 


We start with the Industry applicability criteria.

\textit{Large data volumes and technology stack}

\begin{itemize}
\item  \textit{Scalability}. Authors of Virtual-Taobao \cite{Shi2019} and MARS-Gym \cite{Santana2020}, both dedicated to online retail platforms simulation, state scalability as an important feature for an industrial simulator. Furthermore, MARS-Gym \cite{Santana2020} allows large dataset preprocessing with Apache Spark \cite{10.1145/2934664}, while Virtual-Taobao \cite{Shi2019} reports training from hundreds of millions of real Taobao customers’ records and further successful implementation of RS to real environment. What is more, ReAgent \cite{Gauci2018} is introduced to solve industrially applied RL problems on datasets of millions to billions observations.  The simulator supports CPU, GPU, multi-GPU, multi-node training and data preprocessing with Apache Spark \cite{10.1145/2934664}. In turn, Accordion \cite{McInerney2021} reports the development of a novel scalable algorithm for training models and simulating new trajectories after training.
    
\item \textit{Technology stack}. The majority of considered simulators with available source code are written in Python, except for SynRec \cite{Slokom2018} (R) and DataGenCars \cite{Carmen2017} (Java). Some simulators uses TensorFlow \cite{45381} [RecSim \cite{Ie2019}, RecSim NG \cite{Mladenov2020}]  or PyTorch \cite{10.5555/3454287.3455008} [MARS-Gym \cite{Santana2020}, ReAgent \cite{Gauci2018}] as a computational back-end  which could provide GPU acceleration.
\item \textit{Experiments on large datasets}. Information about datasets used in experiments for each simulator is presented in Table \ref{tab:validating_sims}. Information about the datasets is presented in Table \ref{tab:datasets}. The biggest open datasets used are JD.com [UserSim \cite{Zhao2021}], Netflix Prize [UserSim \cite{Zhao2021}, CF-SFL \cite{Wang2019}] and ContentWise [Accordion \cite{McInerney2021}]. \end{itemize}

\textit{Focus on industry tasks}
\begin{itemize}
\item \textit{Industrial partner}. Industrial partners for all simulators, considered in the survey, are present in Table \ref{tab:existing_sims}. The most notable examples of simulators, impacted industrial RSs are Virtual-Taobao \cite{Shi2019} and RerankSim \cite{Huzhang2021} and ReAgent \cite{Gauci2018}. The authors report the implementation of RSs trained in simulators resulted in the business metrics growth.
 \item \textit{Simulation and real-environment results consistency}. AESim \cite{Gao2021} and RerankSim \cite{Huzhang2021} both show the inconsistency of offline metrics and online performance and the consistency of simulation and A/B tests results. The authors of Accordion \cite{McInerney2021} take user-visit statistics as business metrics and shows a good approximation of A/B test result where user-visit statistics were considered as business metrics. 
 \item \textit{Financial effect}. Virtual-Taobao \cite{Shi2019} and RerankSim \cite{Huzhang2021} report significant performance improvement in the real environment, which leads to a revenue increase.
\end{itemize}

Now let us consider the Solution maturity criteria.
\begin{itemize}
 \item \textit{Entry threshold}. We found detailed documentation, quick-start guides and tutorials for the following simulators: Open Bandit Pipeline \cite{Saito2020}, RecSim \cite{Ie2019}, MARS-Gym \cite{Santana2020}, T-RECS \cite{Lucherini2021}, RecoGym \cite{Rohde2018}, ReAgent \cite{Gauci2018}, RecLab \cite{krauth2020offline}. 
 \item \textit{Open source code with ongoing development and live community}. There are only a few simulators with ongoing development. Two of them are not specialized for RSs [Open Bandit Pipeline \cite{Saito2020},  ReAgent \cite{Gauci2018}]. The other two, T-RECS \cite{Lucherini2021} and RecLab \cite{krauth2020offline},  are specialized for RSs.
\item  \textit{Modular structure and customizability}.  Some simulators state customizability as an advantage in the corresponding research paper, for others we reviewed the source code and found them meeting the criterion, among them: RecSim \cite{Ie2019},  RecSim NG \cite{Mladenov2020}, RecLab \cite{krauth2020offline}, T-RECS \cite{Lucherini2021},  RecoGym \cite{Rohde2018}, MARS-Gym \cite{Santana2020}, ReAgent \cite{Gauci2018}, Open Bandit Pipeline \cite{Saito2020}.
\end{itemize}


Summarizing, only a few simulators, such as Open Bandit Pipeline \cite{Saito2020}, ReAgent \cite{Gauci2018}, MARS-Gym \cite{Santana2020}, T-RECS \cite{Lucherini2021}, RecLab \cite{krauth2020offline} and RecSim \cite{Ie2019} meet the majority of criteria from \textit{Large data volumes and technology stack} and \textit{Solution Maturity Criteria}. It can be noticed that Open Bandit Pipeline \cite{Saito2020} and ReAgent \cite{Gauci2018} are dedicated to general RL tasks with no RS specialization. The others which report valuable business results and meet \textit{Focus on industry task} criteria, are either not publicly available or have too narrow specialization (e.g. designed for a particular business area). Thus, we can conclude that the development of industry-levels simulators with proven business value is in demand by the practitioners.

\section{Variants of implementation of the functional components}

\label{sec:variants}

In this section we overview and discuss particular examples of the functional components {\sf C1}-{\sf C5} defined in  Section~\ref{sec:functionality} for the simulators from Table~\ref{tab:functional_components}.

\subsection{Component {\sf C1} (synthetic data generation)} \label{sec:C1_imp}
In this section we regard the methods for synthetic data generation that are used in the simulators discussed earlier. Note that in addition to the methods described further in this section, there is a huge variety of methods for generating synthetic data  based on various technical ideas and capable of creating datasets of a required type, see e.g. \cite{Patki2016,Dankar2021,Popic2019,Goncalves2020}. The choice of a domain-oriented generative model is usually dictated by a specific recommendation task under consideration.

In general, the choice of user attributes and user response model is challenging. Indeed, it seems that no such model can perfectly capture all the nuances in user behavior \cite{Chaney2021arxiv} and thus one  has to draw on and integrate with various existing models of this type within the simulation of interactions between users and RSs. 

One possible reason for the usage of synthetic data for training and testing RSs is its ability to protect the real data privacy along with the statistical properties \cite{Carmen2017}. Synthetic data should be generated so that the initial real-world dataset must not be re-identified using the synthetic one. In the case of RSs, the most sensitive data is users' preference information and one can use special methods, e.g. from \cite{Slokom2018}, to protect it.

\subsubsection{Models for generating synthetic profiles of users and items}
The authors of RecSim \cite{Ie2019} do not propose a specific model of user and item synthetic profiles generation, but they assume that the attributes are sampled from a prior distribution over the user and item features. Moreover, user's features can vary during the simulation process through a transition model (e.g. user's interest in a particular sphere can increase or decrease with time). RecSim NG \cite{Mladenov2020} does not have a  specification of a model for generation synthetic profiles either but along with the models that are based on sampling features from a prior distribution, RecSim NG contains more complex models, e.g. based on Markov chains and recurrent neural networks.

RecoGym \cite{Rohde2018} assumes the existence of a synthetic user, but its generation process is not described clearly in the corresponding paper. In contrast to RecoGym, Virtual-Taobao \cite{Shi2019} includes a model for generating synthetic profiles of users that is clearly described. The model is based on the generative adversarial network for simulating user feature distributions.

A simple matrix factorization model is used to extract the latent user and item features in ZZA \cite{Zhou2021}. Further, the extracted features serve as synthetic profiles of users and items. T-RECS \cite{Lucherini2021} contains experiments where the profiles of users and items have a synthetic nature (they are sampled from a Dirichlet distribution), but this experiment is the only example of the synthetic data usage in that work.

In DataGenCARS \cite{Carmen2017}, a historical dataset is used to perform statistical analysis of user and item features. The results of the analysis are written in special statistics files that are further converted to a user and item scheme for the synthetic dataset generation. Finally, the authors of SynEvaRec \cite{Provalov2021} propose to use the existing library for synthetic data generation SDV \cite{Patki2016} to create synthetic profiles of users and items with Gaussian Copula, CTGAN and Copula GAN models.

SIREN \cite{bountouridis2019siren} has a {\sf C1}, but the synthetic data provided by it is not based on the real-world one. Users' preferences (or attributes) are sampled from the uniform distribution in the experimental section. As for items' (in this case, they are articles) attributes, they are two-dimensional and sampled from a Gaussian Mixture Model fitted on the BBC document population. The two-dimensional space is the low-dimensional representation of documents that is obtained as a TF-IDF vector of a document and its further t-SNE projection. RecLab \cite{krauth2020offline} provides different implementations of {\sf C1} depending on the environment. Some environments have user and item attributes sampled from a pre-defined distribution. In other environments a factorization model is trained on the historical data, then latent factors representing user's and item's attributes are extracted from it and used as synthetic profiles.

\subsubsection{Models for generating user-item responses}
The authors of RecSim \cite{Ie2019} describe the generation process of user-item responses via a user response model (they do not provide a concrete model). The user response model depends on item's features as well as on user features. In RecSim NG \cite{Mladenov2020} a user response model is based on an affinity model that ingests a set of item features and outputs a vector of scores for a set of items. Then the vector of scores is passed to a distribution that samples a set of chosen items.

In UserSim \cite{Zhao2021}, a user response model is based on a generative adversarial network, where the generator is used to generate synthetic logs based on the historical data, while the discriminator is utilized to predict users' behavior. The generative adversarial network is also used in CLL \cite{Chen2019} for the simulation of user's behavior (in the form of sequential choices) and learning of the user's reward function.

A rather simple user-item response model, that is used for the generation of user feedback on the item that is recommended by an RS (the rating matrix is preliminary debiased), is proposed in SOFA \cite{Huang2020}. A more complex methodology for simulation of user visits, time of the visits and the number of interactions in each visit is proposed in Accordion \cite{McInerney2021}. This methodology is based on Poisson processes and is used for modeling user-item interaction sequences.

The model for generating user-item responses in Virtual-Taobao \cite{Shi2019} is represented by Multi-agent Adversarial Imitation Learning. This model generates users' actions that are further used in the training process. In ZZA \cite{Zhou2021}, the user-item response is modeled by the matrix factorization model, which extracts the latent item features and user preferences from real-word ratings.

In YHT \cite{Yao2021}, the authors analyze several models of synthetic choice and feedback generation. The models are rather simple (such as always positive feedback, feedback based on a threshold of an item's attribute and so on) and are constructed to regard different types of user behavior. The model of generation user responses in CF-SFL \cite{Wang2019} consists of the reward estimator and the feedback generator that is based on a feedback embedding. The whole simulator can be regarded as a method of inverse RL.

The synthetic user-item responses in SynRec \cite{Slokom2018} are generated by a tree, where a Bayesian bootstrap is performed in every node. Note that this method assumes that a part of the historical responses remains the same in the synthetic data. The process of user-item responses generation in the case of DataGenCARS \cite{Carmen2017} is the same as the process of user and item profiles generation: the synthetic dataset (including responses) is generated based on the results of historical data statistical analysis. As for GIDS \cite{Jakomin2018}, its process for generating user-item responses is based on  the assumption that similar users rate similar items in one manner. That is why the first stage of this generator is the division of users and items into clusters. Then, the ratings are sampled according to the probability that a cluster of a particular user is connected to a cluster of a particular item.

A synthetic parametric response function, that can consist of several customizable components (e.g. realistic, heuristic and random) is proposed in SynEvaRec \cite{Provalov2021}. Thus, this framework is able to create parametric classes of synthetic data providing different scenarios of user behavior.

SIREN \cite{bountouridis2019siren} uses a user response model based on multinomial logit, which is used to simulate the choice of articles from a user's awareness pool. The awareness pool is sampled from a complex distribution that can be controlled by a set of parameters. As for RecLab \cite{krauth2020offline}, it has three types of user response model: an explicitly defined function, choices sampled from a defined distribution and choices obtained by a trained on a historical data factorization model.

Let us also mention that \cite{Chaney2021arxiv} provides a survey of other possible user response models including those from economics and marketing. In \cite{Chaney2021arxiv}, the author furthermore promotes the idea that a user response model should take into account that a user may have other offline and online mechanisms for accessing content than RSs (e.g. recommendations from a friend with further searching).

\subsubsection{Modeling scale}

An important issue that is however rarely discussed in the works on simulators is the choice of modeling scale in the user, item and user-item response generation models in {\sf C1}. It is common \cite{Sotomayor2020,Navarro2012} to distinguish micro, meso and macro scales that can be applied for such modeling (possibly in the temporal setting):
\begin{itemize}
    \item at the microscale one models user behavior basing on the individual history of interactions of a chosen user with items (for example, one can approximate the response function for a single user using a long enough history of user-item responses);
    \item at the mesoscale one does it basing on the data for a chosen group (strata) of users, items and the corresponding user-item responses (for example, one can approximate the response function for a group of users with similar features using their group history of user-item responses);
    \item at the macroscale one does it basing on the data for the available population of users and the history of their interactions with items (as in the mesoscale case but for the whole population).
\end{itemize}

Clearly, the modeling scale may influence the possibilities of further scenario modeling in {\sf C2}, for instance, at the mesoscale, one can add a priori assumptions about changes in the preferences of user groups, and at the macro level, take into account external factors that affect the entire population as a whole (for example, the consequences of economical crisis in e-commerce RSs). 

It is common for the simulators in Table~\ref{tab:functional_components} to use the whole population of users to train the generative models for user attributes and user-item responses thus exploiting the macroscale modeling. Some particular simulators e.g. GIDS \cite{Jakomin2018} use for these groups of similar users --- it is an example of the mesoscale modeling. The microscale is not a popular option (probably because of the absence of a long-enough history of user-item interactions for an individual user). Nevertheless, the general-purposed models as in RecSim \cite{Ie2019} may be still related to this scale, at least formally.

As in many other areas of modeling \cite{Sotomayor2020,Navarro2012}, it is reasonable to answer the question about the choice of preferable modeling scale  (or multiscale modeling \cite{Opp2011,Hoekstra2014,Nikhanbayev2019} as an option) for users and their responses on items within a simulator for RSs. It turns out however that this is not answered yet.

\subsection{Component {\sf C2} (scenario modeling)}

In this section, we discuss the methods for scenario modeling that is the base of {\sf C2} in the case of regarded simulators.

As for {\sf C2} in GIDS \cite{Jakomin2018}, it is implemented as an ability to simulate concept drifts in generated data. Moreover, a set of GIDS tunable parameters allows the creation of specific datasets for testing RSs in different scenarios. RecSim \cite{Ie2019} does not contain an implicit component for scenario modeling, but it assumes the ability to test RSs in different regimes, for example, by customizing the level of observability for user and item features.

RecSim NG \cite{Mladenov2020} is probably the most customizable simulator that provides the ability to set parameter values in the behavioral models at different scales (individual or population). The scenario modeling in SynEvaRec \cite{Provalov2021} is provided by a parametric synthetic user-item response function that allows for the evaluation of RSs on data samples with different user behaviors.

YHT \cite{Yao2021} proposes a simple scenario modeling at the level of response models where a special parameter controls the strength of user preference for choosing items with certain attributes. Another type of scenario modeling is the varying degree of preference bias caused by the displayed rating in ZZA \cite{Zhou2021}. The authors introduced the bias as a parameter of a user-item response function, thus, the bias values can be used to model the situations with various degree of displayed rating influence on user choices.

Scenario modeling in SIREN \cite{bountouridis2019siren} simulator is also implemented via a parametric user response model. For example, such parameters as awareness weight and the amount of articles read per iteration per user can be adjusted. Moreover, this simulator allows to control the items attributes via special parameters. RecLab \cite{krauth2020offline} has a similar approach to scenario modeling: user response model can be varied through specific parameters. Moreover, user preferences can be updated according to a parametrically defined function.

It could be noticed that component {\sf C2} of scenario modeling is present in approximately half of the regarded simulators. Moreover, two of the simulators have this component in an implicit form and do not provide any particular implementation. The existing implementations are rather simple and are represented by various parametric response functions. The development of more advanced components of this type seems to be important in future studies.

\subsection{Component {\sf C3} (training and testing RSs)} \label{sec:C3_imp}

In this section, we discuss the variety of datasets and quality metrics often used for training and testing RSs within the component {\sf C4}.

There are many open-source test datasets (listed in Table~\ref{tab:datasets}); their structure and scope are different, however, it is rather difficult to select an open dataset for a specific practical task (especially when data of a certain amount is required for successful model training and testing). The vast majority of open datasets are intended for use in testing RSs not based on RL \cite{Harper2015,Ziegler2005}, while for RL RSs there is much less open data for their analysis \cite{Saito2020,Lefortier2016}. Note that the datasets may be also used for training and testing generative models of user, item and user-item response data. (Recall, however, that despite the variety of open datasets of the real world, their use is associated with certain problems, such as the lack of the possibility of scenario modeling, shifting user preferences, and retraining the RS on a specific dataset.)

The particular datasets, that are used in experiments connected to the simulators under consideration are listed in Table~\ref{tab:validating_sims}. Moreover, this table contains information about the RSs, that are trained and tested in a simulator, and the quality metrics, which are used to evaluate the RSs quality. It is possible to see that RSs based on RL, as well as classic not RL RSs are involved in a simulator pipeline. As for metrics, it should be mentioned that there are standard recommender metrics in Table~\ref{tab:validating_sims} such as MAE, F-measure, etc., that can be computed offline. Moreover, there are specific simulator metrics such as Cumulative reward, CTR, Average CTR, etc., that can be computed only online or in a simulator. One more possible division of metrics used in simulators is into accuracy-based metrics, which compare the recommendation with the user's response (such as Precision, Recall, etc.), and beyond accuracy metrics (such as Novelty, Diversity, etc.). Let us also mention, that the topic of RSs metrics is a well-studied one and, thus, we refer the reader to \cite{Shani2011} for more detailed information. Moreover, Section~\ref{inconsist_metrics} contains some analysis of RSs metrics from the simulation quality control perspective.

To sum up, there is no unified experimental pipeline for training and testing RSs in a simulator. A simulator can be used to train RSs of different types, moreover, this process can be organized by taking into account the requirements of a specific task (e.g. slate recommendations, multiagent simulation). To evaluate the quality of a trained RS, offline metrics based on the results of training  in a simulator can be measured (e.g. \cite{bountouridis2019siren,Shi2019}), as well as online (or business) metrics, which are impossible to measure without a simulator (see \cite{Chen2019,Ie2019}). Furthermore, many open-source datasets are used to compute the metrics values for the RS quality estimation. Besides, real-world datasets are also used for training generative synthetic models for user profiles, item profiles and user-item response functions.

\subsection{Component {\sf C4} (evaluating and controlling simulation quality)}
\label{sec:C4}

This section is devoted to methods used in different simulators for evaluation and control of simulation quality in different senses. As we have mentioned already, one can observe a wide diversity of possible implementations of {\sf C4} as, probably, there is no widely accepted methodology in this direction. For this reason, we also discuss the particular problems related to the simulation usage that motivate the methods.

\subsubsection{Bias and other negative effects studied by simulation}

Let us consider a typical situation in an RS lifecycle. An e-commerce company offers some products to users and exploits an RS for that.
Meanwhile, data scientists in the company work on improving the existing RS. They do offline development as well as evaluation and devise a new RS that outperforms the old one in some offline accuracy-based~\cite{parapar2021} metric like nDCG\footnote{Normalized Discounted Cumulative Gain. This metric takes into account the position of a clicked item in the RS output list, see e.g.~\cite{tamm2021}.}. Then it is a pretty common situation~\cite{Huzhang2021} that once the new RS is deployed in production, the online performance of the new RS is worse than the performance of the old one. This is the example of {\it offline-online inconsistency} so ,in this section, we explore its origins and the solutions simulators may offer to it.
	
There are several reasons for offline-online inconsistency: 
\begin{itemize}
    \item Online environment properties have changed since the data, which has been used for training an RS, was collected;
    \item Unrealistic metrics: accuracy-based metrics, for example, Precision@K and nDCG@K, which are often used in offline RS evaluation do not properly reflect the behavior of a real user;
    \item Negligence of relevant real-world effects: for instance, the current context of a user is not taken into account;
    \item Data biases: the data which is used for training an RS is biased.
\end{itemize}

Let us discuss each one of them in a more detailed manner.

\subsubsection{Online environment change}\label{environment_change}

The online environment may change for many reasons. Some changes are possible to model or predict. For instance, a recommender system itself influences the users' behavior. Also, if there is some steady trend in the online environment like, for example, a steady increase in the number of users, then simulators may try to account for that. Changes that are possible to model or predict are called real-world effects. These effects and the approaches simulators use to address them are discussed in section~\ref{effects_negligence}.

However, some changes are completely unpredictable within the scope of recommender systems and simulators research. These may include
natural disasters, economic crises, and political unrest. All of these may influence the users' behavior drastically. There are also plenty of not-so-prominent effects such as actions of competitors or weather conditions. There is also irreducible noise in the online environment. 

So, clearly, simulators are not able to model all the complexities of the real world and offline-online inconsistencies will always be present. But C4 component of a simulator is intended to reduce those inconsistencies.

\subsubsection{Unrealistic metrics}\label{inconsist_metrics}
To evaluate an RS offline, a typical procedure consists of dividing the available data into training and test sets. The RS is trained on the training set, part of which is usually used for validation. After training, RS is used for making predictions on the test set and comparing predictions with real data using some metrics. These metrics are called accuracy-based since they are computed on a test set using the true responses. The most frequent ones are Precision@K, Recall@K, AUC@K, nDCG@K, MAP@K. However, it has been noticed that the users' behavior is more complex than these metrics would imply. In particular, users would like to discover new content, and value diverse and surprising recommendations. Hence, there are \textit{beyond accuracy} metrics proposed to evaluate an RS output: Novelty, Serendipity, Unexpectedness, Diversity and others~\cite{Silveira2019HowGY}.

 One approach could have been to design a new metric that combines accuracy-based and beyond accuracy metrics~\cite{clarke2008novelty}, ~\cite{parapar2021}. However, this would require tuning the metric for a particular real-world environment. For instance, $\alpha$-parameter in  $\alpha$-nDCG metric~\cite{parapar2021} has to be tuned by human assessors. Therefore, either many A/B tests or many other real-world experiments are required. Moreover, the metric may have to be refitted once users' behavior changes substantially. Hence, none of the simulators, besides RerankSim~\cite{Huzhang2021}, considered in Tables~\ref{tab:functional_components} and~\ref{tab:approbation}, addresses explicitly the issue of metric design.

In a sense, the problem of metric tuning is a \textit{dual} problem to designing a realistic user-response model.
Indeed, as stated above, designing and tuning a new metric is one way to reduce offline-online inconsistency. Another approach is to design a user-response model which reasonably mimics reality. Simulators predominantly follow the second path. Vast majority  of simulators in Table~\ref{tab:functional_components} have a user-response model which is denoted as {\sf C1} component.

To continue, recall that the ultimate goal of a company is to optimize some business metric. They are diverse, depend on domain and business strategy, can be money-related and not~\cite{jannach99}. Typically, business metrics are computed by running a system online. If we want to optimize a real-world business metric offline, we have to have a good model of the real world. Simulators are simplified models of the world which take into account some crucial effects of the real world (see Section~\ref{effects_negligence}). Thus, if properly tuned, simulators may provide a reasonable estimation of (business) metrics without running the online experiments. Simulators that are compared with the real world are usually compared by some business metric like Conversion Rate (CR)~\cite{Huzhang2021}, Rate of Purchase Page (R2P)~\cite{Shi2019}, Number of positive impressions~\cite{McInerney2021}.

The only work which explicitly proposes an ML model which can be used as a metric is RerankSim~\cite{Huzhang2021}. In this work, context is taken into account during training and RL is used to train the model. The model consists of two parts: generator and evaluator. The evaluator's role is to evaluate the quality of the proposed recommendations, it is trained on the historical data. The authors claim that the evaluator model itself serves as a reasonable metric which better approximates the business metrics used in A/B test.

Actually, several other simulators~\cite{Chen2019},\cite{Zhao2021} which are designed for RL RSs demonstrate that reward approximation by RL model could be a better approximation of real-world performance than an offline metric. This leads to a discussion on what should be called a metric. Should it be computed by an explicit formula or an ML model can also be used as a metric? Nevertheless, there are signs that a score computed by a simulator or an RL RS itself better reflects a real-world performance than conventional offline metrics.

\subsubsection{Real-world effects negligence}\label{effects_negligence} 

Simulators are simplified models of the real world in which an RS operates. However, some effects can be taken into account, thus making the simulators closer to the real world.

The main effect which is implemented by most of the simulators is sequential interaction with an RS. Its other name is an RS feedback loop since the recommendations provided by the RS are converted to user interactions which in turn are used for the (partial) RS update.
Next, in Accordion simulator~\cite{McInerney2021} the effect of recurring users' visits is considered. If a user likes recommendations he visits the service more often, and on the opposite, he leaves the service if the recommendations are not relevant. The RerankSim~\cite{Huzhang2021}, model takes into account the whole slate of recommendations (context) since user choice is affected by all items in a slate. The authors of DataGenCars~\cite{Carmen2017} also consider the context in their model. Another effect that is considered in several simulators is the users' \textit{preference drift}. That is, users' preferences change with time under the influence of various factors. The simulators which model it are: RecoGym~\cite{Rohde2018}, Reclab~\cite{krauth2020offline}, SIREN~\cite{bountouridis2019siren}, GIDS~\cite{Jakomin2018}.  

The more real-world effects are taken into account, the more accurate the simulator may be, but, at the same time, it is more difficult to tune it. Currently, the issue of tuning the simulator or its hyper-parameters is not systematically discussed in the literature.

\subsubsection{Data biases}\label{data_biases}

\begin{figure}
\centering
    \includegraphics[width=0.7\linewidth]{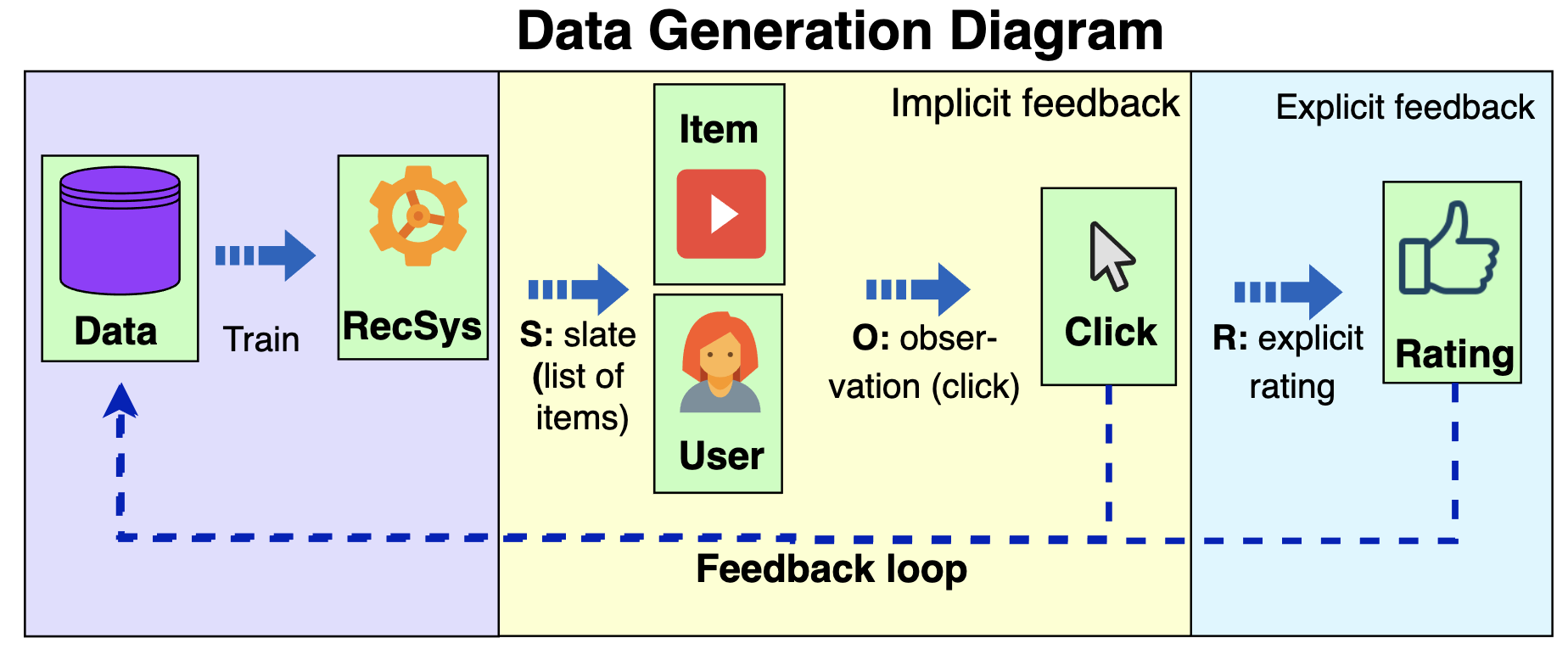}
    \caption{Data generation process. It facilitates better understanding of data biases}
    \label{fig:recsys_data_generation}
\end{figure}

\begin{table}
  \caption{Biases in user-items interactions, their origins and descriptions}
  \label{tab:biases}
  \centering
  \begin{tabular}{p{3cm}|p{5.5cm}|p{2.5cm}|p{2cm}}
    \hline
    \textit{Entity characteristics} & \textit{Related bias and its description} & \textit{Synonyms (from ~\cite{Chen2020Bias})} & \textit{Explicitly studied in simulators} \\
    \hline
    \hline
     \textit{User} &  &  & \\
    \hline
    \textit{True item preference}. It may change in time. It can also be stochastic e.g. depend on user's mood and environment & \textit{Positivity bias}: users tend to evaluate/click items they like more & \textit{Selection bias}: has more general definition i.e. data is biased because users have freedom to choose items & SOFA \cite{Huang2020} \\
    \textit{Community of a user} & \textit{Conformity bias}: users tend to agree with their peers/influences in item evaluation & & \\
    \textit{Socio-demographic characteristics} & \textit{Unfairness}: users are unevenly distributed in the datasets. Some groups may be underrepresented & & \\
    \hline
    \hline
    \textit{Item} & & & \\
    \hline
    \textit{Item popularity} & \textit{Popularity bias}: users tend to interact with more popular items & & SOFA \cite{Huang2020}, YHT \cite{Yao2021}  \\
    \hline
    \hline
    \textit{RS} & & & \\
    \hline
    {\it Recommended slate} & \textit{Exposure bias}: users see predominantly recommended items so they are not aware of other items. Matthew effect and filter bubbles are related & \textit{Previous model bias, User-selection bias} & \\
     & \textit{Preference bias}: slate's content and order influence user's clicks/ratings. Decoy effect~\cite{Huzhang2021} is related & \textit{Position bias} & RerankSim \cite{Huzhang2021}, ZZA \cite{Zhou2021} \\
  \end{tabular}
\end{table}

Another reason for the offline-online inconsistency is biases present in the user and interaction data. The cause of data biases is that the data we receive is most of the time observational rather than experimental. Hence, the collected data is subject to all imbalances present in the real-world data collection process. A recent overview of biases is provided in~\cite{Chen2020Bias}.

We summarize biases in a slightly different way which is based on a data generation process shown in Figure~\ref{fig:recsys_data_generation}. The diagram shows the core of data generation, however, in different applications there can be other details. The data generation starts from some baseline RS which recommends a slate of items to a user. Note, that in reality, a user can interact not only with items RS recommends, she can find items independently, but this detail is not substantial here. Then a user selects an item and clicks on it. If we are in the implicit response setup, then the interaction ends here. In case we are in the explicit setup a user evaluates an item after consuming it. In the unbiased case, the click (or observation) would happen independently of a user, item, and the slate. In practice, because observations are often suggested in the slate and users as well as items have their characteristics, the observations and ratings depend on those characteristics and the data becomes biased. 

The description of biases and the cause of their origins are presented in Table~\ref{tab:biases}.  In the leftmost column of Table~\ref{tab:biases}, these intrinsic characteristics are listed. In the second column, we describe biases that appear because these internal characteristics influence observations. In the third column, we give biases' synonyms and variations. For instance, one assumption is that a user has his own intrinsic preference for each item and the goal of an RS is to unveil it. However, users prefer to interact with items they like more, so in a collected dataset, items that a user does not like are underrepresented. This is a \textit{positivity bias}. Since users have other features like their friends and influences, users' clicks/ratings are biased towards their community. \textit{Popularity bias} is frequently observed, users tend to interact with more popular items and as a result, popular items are overrepresented in the data even more than their popularity would suggest. 
\textit{Exposure bias} happens because a user sees the recommended slate and is not aware of plenty of other items. \textit{Preference bias} is a bias caused by the order of items in a slate as well as the content of a slate since the selection of an item may depend on other items in a slate (decoy effect)~\cite{Huzhang2021}.

There is usually a feedback loop in the RS usage, that is, an RS influences which items a user will interact with on the next step. These interactions, in turn, are added to the collected data. The feedback loop may create such phenomenons as ``filter bubbles''~\cite{nguyen2014} , ``rich-get-richer effect''~\cite{fleder2009} ``echo chambers''~\cite{ge2020}, ``Matthew effect''~\cite{merton1968} and so forth, so the biases may amplify. It is worth saying, that the terminology of biases in the RS literature has not been settled yet. So, the same biases may be called differently in different literature and different terms may mean the same. Below we disambiguate this terminology. Disproportions of users in gender, age, race and other socio-demographic characteristics in the data are called unfairness, however, essentially, it is also a type of data bias~\cite{Chen2020Bias}.

According to our analysis, the data bias removal problem is addressed only in one simulator, SOFA \cite{Huang2020}, while in YHT \cite{Yao2021} and ZZA~\cite{Zhou2021} the impact of bias is only analyzed.
In contrast, RS literature has a decent set of methods of bias removal methods, their review is given in~\cite{Chen2020Bias}. Identification and removal of biases in simulators is an open direction for further reducing offline-online inconsistencies.

\subsubsection{Consistency of simulation results on synthetic and real-world data}

One (optional) procedure that may be attributed to the component  {\sf C4} (and particularly to {\sf C1}) is the quality evaluation of a generative model for synthetic user, item and user-response data. This procedure can be implemented in several ways. The first one is to evaluate the synthetic data quality by a set of metrics (e.g. one can use those from the SDV library \cite{Patki2016}). These metrics are divided into groups: statistical metrics (compare the data distributions by statistical tests), likelihood metrics (evaluate the likelihood of the synthetic data on a fitted to the real data model), detection metrics (evaluate how it is easy to distinguish real and synthetic data by a model), machine learning efficacy metrics (evaluate the performance of a trained on a synthetic data model on a real data),  privacy metrics (evaluate the probability of prediction real data attributes given the synthetic data).

Another approach from {\sf C4} for quality evaluation of the generative models for synthetic data is the analysis of how RSs under comparison work on real-world data and the corresponding synthetic one (thus the data from {\sf C1} and {\sf C3} are involved in the process). It is reasonable to expect within the simulation that the RS comparison results are consistent e.g. in the sense that the list of RSs ranked by quality on real-world data is the same as for the synthetic one. However, this is not widely discussed in papers on simulators and moreover, some of the results look contradictory. For example, the authors of SynRec \cite{Slokom2018} state that their results are consistent in the above-mentioned sense. However, in the case of SynRec \cite{Slokom2018} the metric values are almost vanishing and the paper does not have the deviation analysis over several runs, as we have noticed. In our opinion, this issue requires additional analysis. From other studies, for example, DataGenCARS \cite{Carmen2017}, it is however seen that the RS comparison results may be not consistent.

Definitely, this may be considered as an essential drawback of simulator studies as make the practical use of simulators with inconsistent results of RS comparison on real-world and synthetic data questionable.

Let us also mention that one specific implementation of {\sf C4} is presented in Virtual-Taobao \cite{Shi2019}, where a special strategy is proposed to reduce overfitting of an RS in the simulator. This is an attempt to overcome offline-online inconsistency, as this strategy controls the RL algorithm actions, balancing between the accuracy on the historical data and the improvement within the simulated environment.

Finally, to make conclusions about the simulation quality, one can also compare the values of business metrics computed within the simulation and those obtained as a result of online testing during the implementation of a new RS (see e.g. how it is done in Accordion \cite{McInerney2021}). 

\subsection{Component {\sf C5} (summarizing experimental results)} \label{sec:C5_imp}
This section contains the description of methods for summarization of experimental results that are provided in the regarded simulators. The variants of {\sf C5} implementation can be classified according to a component whose results are summarized. For example, a user response model (belonging to {\sf C1}) and its quality comparison with the same models of other simulators is visualized in UserSim \cite{Zhao2021}. Another possible option related to {\sf C1} is a visualization of user preferences drift over time in the topic space in SIREN \cite{bountouridis2019siren}.

As for {\sf C2}, the results of scenario modeling are presented in SynEvaRec \cite{Provalov2021} in the form of a discrete approximation of the ``best quality surface'' for a chosen dataset and a set of RSs. Namely, one can see in the plot the best quality result for each set of scenario parameters used in the parametric response function. Moreover, {\sf C2} results are visualized as the changes in item popularity during simulation for different types of user behavior (defined as a parametric function), see YHT \cite{Yao2021} and ZZA \cite{Zhou2021}.

The learning process taking place in {\sf C3} is usually presented as a plot showing the behavior of RSs quality metric over training steps: for example, in CLL \cite{Chen2019}, CF-SFL \cite{Wang2019}, \cite{Shi2019} (improvement of RL-based RS to supervised learning one in terms of two metrics over time), ZZA \cite{Zhou2021}, SIREN \cite{bountouridis2019siren} (the behavior of two diversity metrics), RecLab \cite{krauth2020offline} (with respect to the mean rating, also). Moreover, tables with the results of RSs quality comparison in terms of various metrics can be found in CLL \cite{Chen2019}, CF-SFL \cite{Wang2019}. RecLab \cite{krauth2020offline} contains figures showing the dependence between an RS quality metric and the mean user ratings of recommended items in a selected environment for a set of RSs.

Finally, {\sf C5} is presented in SOFA \cite{Huang2020} as a tool for bias effects visualization. There are histograms showing rating distributions generated with and without a debiasing method and plots presenting a comparison of learning curves, which track an online metric over time, for policies learned in a simulator with and without a debiasing step. 
Authors come to the conclusion that the policies resulting from using the simulator with the proposed debiasing procedure outperform the policy resulting from using the simulator without this step.

As for special technical instruments, that can be related to {\sf C5}, it should be mentioned that the authors of T-RECS \cite{Lucherini2021} develop a tool for tracking the results of experiments. It is possible to specify the desired metric, and the simulator will count its values at each iteration.

To conclude, there are different implementations of {\sf C5} that are based on visualization of various kinds. However, there is no simulator that provides an automatic summarization of the obtained simulation results, e.g. an automatic inferring of a set of parameters that are optimal for a user (a group of users) with respect to a chosen RS quality metric. Here, under a set of parameters, we mean synthetic generation models with their parameters, response model with its parameters, parameters of a scenario modeling, RS with its parameters and simulation quality control instruments. Clearly, such tools may be helpful in the analysis of extensive experimental results.

\bigskip

\section{Conclusions}

The analysis of scientific works in the area under consideration allows us to conclude that M\&S of interactions between users and RSs implemented in the form of synthetic data-based simulators:
\begin{itemize}
\item have a great potential for accelerating the research and industrial deployment of RSs;
\item provide a compromise (in terms of complexity of implementation, controllability and compliance with reality) between systems for Counterfactual Police Evaluation and Online Controlled Experiments;
\item help to preserve the privacy of and supplement/replace  real-world data by using its synthetic analogues;
\item give the opportunity to train and test RSs under different what-if scenarios;
\item are useful for studying and overcoming negative effects of long-term interactions between users and RSs.
\end{itemize}
From one side, these factors have motivated the development of simulators by multiple teams of researchers and practitioners, including those from well-known companies such as Google, Facebook, Baidu, Netflix and Alibaba that actively use RSs as a part of their e-commerce systems. Another side of the interest towards the M\&S in both academic and industrial spheres is that now there exists a variety of theoretical motivations, approaches, and implementations  resulting in the absence of consensus on best practices in the field.

To clarify the situation, in the current paper we provided  a comprehensive
overview of the recent works on simulators, evaluated and compared them by means of a new consistent classification criteria. Within the classification, we distinguished simulators with respect to their functionality (the presence of functional components), approbation (the reproducibility of the simulator's experimental study) and industry effectiveness (the suitability for industrial deployment). 
Furthermore, we analyzed existing variants of implementation of the simulator's functional components:  {\sf C1} (generating synthetic data),  {\sf C2} (scenario modeling), {\sf C3} (training and testing RSs), {\sf C4} (evaluating and controlling simulation quality) and {\sf C5} (summarizing experimental results). In these terms one can determine the following emerging topics in the recent research on simulators:
\begin{itemize}
    \item improvement of synthetic data generation quality by taking into account additional real-world effects (in particular, such as different types of bias in the historical data) in {\sf C1};
\item development of new simulators with the more advanced implementation of {\sf C2} and {\sf C3} allowing for modeling e.g. the impact of user preference drift, bias in the historical data and long-term negative effects of interactions between users and RSs on the performance of RSs;
\item  attempts to develop and improve methods for simulation quality evaluation and control in {\sf C4} (and partly in {\sf C5}) and to close the simulation-to-reality gap by different means;
\item increase the reproducibility of simulators by means of the public availability of source code with detailed documentation, datasets and experimental results;
\item development of more flexible and customized simulators.
\end{itemize}

Nevertheless, in spite of the active research in the area, we find it important to state the following open problems discovered within our survey:
\begin{itemize}
\item the inconsistency of RS comparison results on real-world and synthetic data (although it is reasonable to expect the opposite) and, in general, the lack of comparative analysis of methods for assessing the quality of and for generation of synthetic data most suitable for training and testing RSs;
\item the complexity of assessing and controlling the quality of simulators and the lack of universally recognized methodology for it\footnote{Potentially the existing fundamental approaches from other topic areas (see e.g. \cite{VanHorn1971,Sargent2013,Balci1995}) may be useful for this purpose.};
\item the lack of extensive experimental comparison (under the same settings such as datasets, quality metrics and RSs under consideration) of existing simulators and their components;
\item the shortage of methodologically solid and reproducible results that quantitatively and qualitatively approve the practical usefulness of simulators (e.g. quality or revenue increase in the real environment).
\end{itemize}

We hope that the observations made in the current paper and the statement of the above-mentioned open problems will motivate further focused research in the field of M\&S of interactions between users and RSs and applications of the M\&S to the performance improvement of industrial recommender engines.

\section*{Declaration of interests}

The authors declare that they have no known competing financial interests or personal relationships that could have appeared to influence the work reported in this paper.

\section*{Author contributions}

Elizaveta Stavinova: Methodology; Visualization; Writing --- original draft; Writing --- review \& editing; Alexander Grigorievskiy: Conceptualization; Methodology; Visualization; Writing --- original draft; Writing --- review \& editing; 
Anna Volodkevich: Conceptualization; Methodology; Writing --- original draft; Writing --- review \& editing; Petr Chunaev: Conceptualization; Methodology; Project administration; Visualization; Writing --- original draft; Writing --- review \& editing; Klavdiya Bochenina: Project administration; Writing --- review \& editing; Dmitry Bugaychenko: Conceptualization, Writing --- review \& editing.

\bibliographystyle{elsarticle-harv}
 \bibliography{SimulatorCite}

\appendix
\newpage
\section{Additional tables}
\label{sec:sample:appendix}

{
\scriptsize
\begin{longtable}{p{2cm}|p{2.7cm}|p{2.1cm}|p{2.7cm}|p{2.7cm}}
\caption{Brief overview of existing simulators}
\label{tab:existing_sims}\\
Simulator & Brief description & Industrial affiliation  & Experiments & Conclusions\\
\hline
DataGen CARS \cite{Carmen2017} & A synthetic dataset generator for the evaluation of context-aware RSs. & \xmark & Comparison of ratings distributions in real-world and synthetic data; comparison of RSs performance on real-world and synthetic data. & The percentage of real-world and synthetic user ratings are nearly identical; the absence of preservation of the quality hierarchy of RSs on real-world and synthetic data. \\
\hline
GIDS \cite{Jakomin2018} & A generator of inter-dependent data streams capable of generating temporal synthetic datasets for RSs. & \xmark & Comparison of statistics distributions in real-world and synthetic data; comparison of RSs performance on real-world and synthetic data. & The distributions of real-world and synthetic data statistics are close; preservation of the quality hierarchy of RSs on real-world and synthetic data. \\
\hline
RecoGym \cite{Rohde2018} & An RL environment for recommendation, where a user is supposed to have two types of interactions with the items. & Criteo & \xmark & \xmark \\
\hline
SynRec \cite{Slokom2018,Slokom2020arxiv} & A framework that uses data synthesis for RSs comparison. & \xmark & Comparison of ratings distributions in real-world and synthetic data; comparison of RSs performance on real-world and synthetic data. & The percentage of real-world and synthetic user ratings are nearly identical; preservation of the quality hierarchy of RSs on real-world and synthetic data.\\
\hline
SIREN \cite{bountouridis2019siren} & A simulation framework that allows to analyze different recommenders' performance with respect to two metrics. & \xmark & Comparison of RSs performance in terms of two diversity metrics over simulation iterations. & Conclusions about the best performing algorithms with respect to the chosen metrics. \\
\hline
CLL \cite{Chen2019} & An RL-based framework for RSs, where a GAN is used to imitate user behavior dynamics. & \xmark & The predictive accuracy of GAN user model is assessed; a recommender police is derived with a learned user model. & The GAN user model is able to capture user interest evolution; the policy based on GAN user model results in higher reward for most users comparing to baseline methods. \\
\hline
RecSim \cite{Ie2019} & A platform for creation of a simulation environments for RSs that supports sequential interaction with users. & Google & Agent strategies comparison (in terms of CTR). & \xmark \\
\hline
Virtual-Taobao \cite{Shi2019} & A simulator learned from historical user behavior data for policies training. & Alibaba & Comparison of the traditional supervised learning approach and the recommender strategy trained in Virtual-Taobao. & The strategy trained in Virtual-Taobao achieves more than 2\% improvement of revenue in the real environment. \\
\hline
CF-SFL \cite{Wang2019} & A framework that improves collaborative filtering with a synthetic feedback loop. & Facebook & Performance comparison between the CF-SFL framework and various RS baselines; learning trajectories comparison of the mentioned approaches. & Improvements over the baselines on all evaluation metrics with the proposed CF-SFL framework; the CF-SFL framework can improve the performance after the pre-training steps (contrary to VAE whose performance stays the same). \\
\hline
SOFA \cite{Huang2020} & A simulator that accounts for interaction biases present in historical data prior to optimization and evaluation. & \xmark & Quality analysis (in terms of cumulative number of clicks) of policy trained using a debiased simulator and a naive simulator. & Policies trained on a debiased simulator perform better than the policies resulting from using naive simulator. \\
\hline
RecLab \cite{krauth2020offline} & A simulation framework for evaluation of recommenders across simulated environments. & \xmark & Comparison of RSs online performance, RSs offline metrics and their relationship. & Offline metrics are correlated with online performance over a range of environments. \\
\hline
RecSim NG \cite{Mladenov2020} & A probabilistic platform for the simulation of multi-agent RSs that supports sequential interaction with users. & Google & The dependence between the user's utility and item's provider policies. & The longer the user's history, the more reward the system can obtain. \\
\hline
T-RECS \cite{Lucherini2021} & A Python package developed to simulate RSs and other types of sociotechnical systems to study their societal impact. & \xmark & Conceptual replication of several simulation-based studies using the T-RECS. & The replications demonstrate the power of T-RECS to carry out complex algorithmic simulations. \\
\hline
Accordion \cite{McInerney2021} & A simulator based on Poisson processes that can model temporal visit patterns to a system. & Netflix & Comparison of the proposed simulator and the debiasing method Norm-IPS in the task of predicting the results of A/B testing. & Accordion's forecast is more accurate. \\
\hline
SynEvaRec \cite{Provalov2021} & A framework for evaluating and comparing RSs using synthetic user and item data and parametric user-item response functions. & \xmark & Comparison of RSs (SVD, kNN, NMF)  under different scenarios of users behavior. & Quality surfaces indicating the RS showing the best quality under a certain combination of response function parameters. \\
\hline
YHT \cite{Yao2021} & A simulation framework for measuring the impact of an RS under different types of user behavior. & Google & Analysis of user trajectories under different models of user behavior and two RSs (MF-based and NN-based). & RSs show non-trivial temporal dynamics with every type of user behavior. \\
\hline
UserSim \cite{Zhao2021} & A simulator of a user behavior (actions and responses) based on a GAN. & Baidu & Similarity analysis of the initial sequences of item-response pairs of each user interactions and the sequences generated by different methods; simulators comparison in terms of user response prediction. & UserSim shows the best performance (in terms of ROUGE-1) among the methods for sequences generation, as well as the best performance (in terms of F1, AUC) among the simulators. \\
\hline
ZZA \cite{Zhou2021} & A simulation approach to study the impact of preference biases on the RSs performance. & \xmark & The quality of two RSs (SVD and kNN) at different preference biases is analyzed. & A high preference bias results in lower recommendation accuracy and relevance; when the preferences of half of the users are biased, the recommendation accuracy is in between no-bias and high-bias populations, while the consumption relevance is close to the no-bias population. \\
\end{longtable}
}

\newpage

{\scriptsize
\begin{longtable}{p{2cm}|p{11cm}|p{0.5cm}}
\caption{Open datasets that can be used in simulators}
\label{tab:datasets}\\
Dataset & Description & Availability\\
\hline
LDOS-CoMoDa & 1600 movie ratings, more than 90 users with attributes (age, gender, city, country), 950 movies with attributes (director, country, language, year, etc.). & \href{https://www.lucami.org/en/research/ldos-comoda-dataset/}{Link} (by request) \\
\hline MovieLens (100K as an example) &
100 thousand movie ratings, 943 users with attributes (age, gender, occupation), 1682 films. & \href{https://grouplens.org/datasets/movielens/100k/}{Link} \\
\hline
Yahoo Music (R3, version 1.0 as an example) &
311 thousand ratings for songs, 15 thousand users, 1,000 songs. &	\href{https://webscope.sandbox.yahoo.com/catalog.php?datatype=r}{Link} \\
\hline
Yelp &
6.9 million reviews (including ratings) for different businesses, 1.9 million users, 150 thousand businesses with attributes (opening hours, parking, accessibility, surroundings, etc.). & \href{https://www.yelp.com/dataset}{Link} \\
\hline
GoodBooks 10K &
982 thousand ratings for books, 53 thousand users, 10 thousand books with attributes (author, year, title, etc.). &	\href{https://www.kaggle.com/code/philippsp/book-recommender-collaborative-filtering-shiny/data}{Link} \\
\hline
LastFM (360K as an example) &
17 million listening records (number of plays), 359 thousand users, 300 thousand artists. &	\href{https://www.upf.edu/web/mtg/lastfm360k}{Link} \\
\hline
Taobao & 100 million records about user behaviors (implicit feedback that has four categories), 987 thousand users, 4 million items. & \href{https://tianchi.aliyun.com/dataset/dataDetail?dataId=649}{Link} \\
\hline
YooChoose & 33 million of user clicks and 1 million of user buys in an e-commerce platform, that form 9 million unique user sessions. & \href{https://www.kaggle.com/datasets/chadgostopp/recsys-challenge-2015}{Link} \\
\hline
Netflix Prize &
100 million movie ratings, 480 thousand users, 18 thousand movies (with a title and a year).	& \href{https://www.kaggle.com/netflix-inc/netflix-prize-data}{Link}\\
\hline
Million Song & 33 million ratings for songs, 571 thousand users, 41 thousand songs with attributes (artist, tags). & \href{http://millionsongdataset.com/}{Link} \\
\hline
Coat & Ratings from 290 users on 24 self-selected items and 16 randomly-selected items from totaly 300 items in a web-shop. & \href{https://www.cs.cornell.edu/~schnabts/mnar/}{Link} \\
\hline
ContentWise impressions & 10 million interactions between users of a media service and media content, 42 thousand users, 145 thousand items. Moreover, the dataset contains 23 million of impressions (the recommended items that were presented to the user). & \href{https://github.com/ContentWise/contentwise-impressions}{Link} \\
\hline
Restaurants and Their Clients & 1 thousand ratings for restaurants, 138 users with attributes (smoker, drink level, dress preference, etc.), 769 restaurants with attributes (the availability of a certain cuisine type). & \href{https://www.kaggle.com/datasets/uciml/restaurant-data-with-consumer-ratings}{Link} \\
\hline
Book-Crossing &
1.1 million ratings for books, 278 thousand users (anonymized, but with age and address), 271 thousand books. &	\href{https://www.kaggle.com/datasets/arashnic/book-recommendation-dataset}{Link} \\
\hline
JD.com & 633 thousand trajectories (with feedback) of users' accessing logs from an e-commerce platform, 471 thousand users with profiles (dimension of 20), 456 thousand items. & \href{https://datascience.jd.com/page/opendataset.html}{Link} (by request) \\
\hline
Amazon Review Data (2018 as an example) &
233 million ratings and reviews for products (with information such as color, size, package type, etc.), no user attributes. & \href{https://nijianmo.github.io/amazon/index.html}{Link} \\
\hline
Jester (Dataset 1 as an example) &
4.1 million ratings for jokes, 73 thousand users with no attributes (only the number of rated jokes), 100 jokes with a text description.  & \href{https://eigentaste.berkeley.edu/dataset/}{Link} \\
\hline
Open Bandit Dataset &
26 million records of users clicks in an online clothing store, each record is a user reaction with a timestamp, characteristics of the user, item, location on the site, probability of recommendation with the existing strategy &	\href{https://research.zozo.com/data.html}{Link} \\
\hline
Criteo Data &
103 million records made by an online advertising company, each record contains the characteristics of the user, the advertisement shown to the user (i.e. agent actions), click markers (i.e. reward), the probability of showing one or another ad to the user. Suitable for bandit problems. &	\href{https://ailab.criteo.com/download-criteo-1tb-click-logs-dataset/}{Link} \\
\hline
Yahoo News (R6A as an example) &
45 million records of user clicks on a new site, each record contains user characteristics, featured news (i.e. actions), click indicators (i.e. rewards). Suitable for bandit score. & \href{https://webscope.sandbox.yahoo.com/catalog.php?datatype=r}{Link} \\
\hline
RL4RS &
2 million user sessions on the online game platform, 156 thousand users, 381 items Each record contains items shown to the user, his/her reaction, characteristics of the user and items, timestamp. & \href{https://drive.google.com/file/d/1YbPtPyYrMvMGOuqD4oHvK0epDtEhEb9v/view}{Link} \\
\hline
Finn.no Slates &
37.4 million user on the Norwegian ad site interactions,  2.3 million users, 1.3 million items with attributes. Itemss were shown to the user either as a result of a query or as a result of a recommendation algorithm. & \href{https://github.com/finn-no/recsys_slates_dataset}{Link} \\
\end{longtable}
}

\newpage

{\scriptsize
\begin{longtable}{p{3cm}|p{3cm}|p{3cm}|p{3cm}}
\caption{Datasets, RSs and metrics used in simulators}
\label{tab:validating_sims}\\
Simulator & Dataset & RSs & Metrics \\
\hline
DataGenCARS \cite{Carmen2017}  &
LDOS-CoMoDa & CM (Contextual Modeling), SVD & MAE, F-measure \\
\hline
GIDS

\cite{Jakomin2018} &
MovieLens, Yahoo Music, Yelp & RMF, NMF, PMF, etc. & RMSE, RRMSE, MAE, F1, P@10, R@10 \\
\hline
RecoGym

\cite{Rohde2018} & Open real-world data is not used & –-- & --- \\
\hline
SynRec

\cite{Slokom2018,Slokom2020arxiv} &
MovieLens, GoodBooks  & MF, BMF, BPRFM & Recall@5 \\
\hline
SIREN 

\cite{bountouridis2019siren} &
Open real-world data is not used & ItemKNN, UserKNN, MostPopular, WeightedBPRMF & EPC (Expected Popularity Complement), EPD (Expected Profile Distance) \\
\hline
CLL

\cite{Chen2019} & MovieLens, Last.fm, Yelp, Taobao, YooChoose & W\&D-LR, W\&D-CCF, GAN-Greedy, etc. & Cumulative reward, CTR \\
\hline
RecSim

\cite{Ie2019} &
Open real-world data is not used & TabularQAgent, FullSlateQAgent, UCB1 & Average CTR \\
\hline
Virtual Taobao \cite{Shi2019} &
Open real-world data is not used &
TRPO, regression, regularized regression & Total Turnover, Total Volume, Rate of Purchase Page \\
\hline
CF-SFL

\cite{Wang2019} & MovieLens, Netflix Prize, Million Song & SLIM, WMF, CDAE, etc. & Recall@20, Recall@50, NDCG@100\\
\hline
SOFA 

\cite{Huang2020} &
Yahoo Music, Coat & DQN & Cumulative number of clicks, MSE, MAE, etc. \\
\hline
RecLab

\cite{krauth2020offline} & MovieLens & TopPop, ItemKNN, UserKNN, Oracle, SGD MF, Bayes MF, AutoRec, CF-NADE, LLORMA, EASE & RMSE, NDCG \\
\hline
RecSim NG \cite{Mladenov2020} &
Open real-world data is not used & Simple reinforce policy, several manually constructed policies & Average Cumulative Reward, Cumulative User Utility \\
\hline
T-RECS \cite{Lucherini2021} & Open real-world data is not used & Popularity Recommender, Content Filtering, Social Filtering, etc. & Measurement of Structural Virality, Interaction Measurement, Interaction Spread, etc. \\
\hline
Accordion

\cite{McInerney2021} & ContentWise impressions & NMF & --- \\
\hline
SynEvaRec \cite{Provalov2021} & Restaurants and Their Clients, Book-Crossing & kNN, NMF, SVD & RMSE \\
\hline
YHT

\cite{Yao2021} & MovieLens & MF, RNN & Item Popularity, Slope of a user trajectory \\
\hline
UserSim

\cite{Zhao2021} &
JD.com & DQN & Average Reward\\
\hline
ZZA 

\cite{Zhou2021} &
Netflix Prize & SVD, UserKNN, ItemKNN & RMSE, Relevance, Diversity\\
\end{longtable}
}

\newpage





\end{document}